\documentclass[journal]{IEEEtran}
\IEEEoverridecommandlockouts
\usepackage{graphicx}

\usepackage{cite}
\usepackage{url}
\usepackage{mdwmath}  
\usepackage{soul}
\usepackage[table]{xcolor}
\usepackage{amsfonts,amsmath,amssymb,amsxtra,amsbsy,floatflt}
\usepackage{indentfirst} 
\usepackage{tabularx}
\usepackage{multirow}
\usepackage{nth}
\usepackage{comment}
\usepackage{epstopdf}
\usepackage{hhline}
\usepackage{siunitx}
\usepackage{booktabs}
\usepackage[bottom]{footmisc}
\usepackage[acronym]{glossaries}
\usepackage{upgreek}
\usepackage{algorithm}
\usepackage{algorithmicx,algpseudocode}
\usepackage{bm}
\usepackage{array} 
\usepackage{epsfig}
\usepackage{times}
\usepackage{framed}
\usepackage{enumitem}
\usepackage{float}
\usepackage{algpseudocode}
\usepackage{mathtools}
\usepackage{bbm}
\usepackage{makecell}
\usepackage{pifont}
\usepackage[export]{adjustbox} 
\usepackage{latexsym}
\usepackage{multicol}
\usepackage{eurosym}
\usepackage{pdfpages}
\usepackage{verbatim}
\usepackage{stfloats}

\usepackage{mathtools}
\usepackage{algpseudocode}
\usepackage{amssymb}
\usepackage{amsmath}

\def\1{\mathbf{1}}

\newcounter{subeqn} 
\makeatletter
\@addtoreset{subeqn}{equation}
\makeatother

\newcommand{\name}{RiLoCo}

\newtheorem{problem}{Problem}
\newtheorem{theorem}{Theorem}
\newenvironment{sketch}{{\noindent \it Sketch of Proof:~}}

\newcommand{\change}[1]{{\color{black} {#1}}}

\newacronym{3d}{3D}{three dimensional}
\newacronym{bs}{BS}{base station}
\newacronym{ue}{UE}{user equipment}
\newacronym{roi}{ROI}{return of investment}
\newacronym{cs}{CS}{Candidate Site}
\newacronym{rl}{RL}{reinforcement learning}
\newacronym{nn}{NN}{neural network}
\newacronym{ml}{ML}{machine learning}
\newacronym{ehf}{EHF}{extremely high frequency}
\newacronym{crb}{CRB}{Cramér–Rao bound}
\newacronym{toa}{ToA}{time of arrival}
\newacronym{pdf}{PDF}{probability density function}
\newacronym{csi}{CSI}{channel state information}
\newacronym{aoa}{AoA}{angle of arrival}
\newacronym{ula}{ULA}{uniform linear array}
\newacronym{ura}{URA}{uniform rectangular array}
\newacronym{uca}{UCA}{uniform circular array}
\newacronym{mimo}{MIMO}{multiple-input, multiple-output}
\newacronym{los}{LoS}{line-of-sight}
\newacronym{nlos}{NLoS}{non-line-of-sight}
\newacronym{b5g}{B5G}{beyond fifth-generation}
\newacronym{6g}{6G}{sixth-generation}
\newacronym{snr}{SNR}{signal-to-noise ratio}
\newacronym[plural=RISs, firstplural=reconfigurable intelligent surfaces (RISs)]{ris}{RIS}{reconfigurable intelligent surface}
\newacronym{kpi}{KPI}{key performance indicator}
\newacronym{isac}{ISAC}{Integrated Sensing and Communication}

\begin{document}

\title{\name: An ISAC-oriented AI Solution \\to Build RIS-empowered Networks}

\author{Guillermo Encinas-Lago,~\IEEEmembership{Student Member,~IEEE,}
Vincenzo~Sciancalepore,~\IEEEmembership{Senior Member,~IEEE,}\\
Henk Wymeersch,~\IEEEmembership{Fellow,~IEEE,} Marco Di Renzo,~\IEEEmembership{Fellow,~IEEE,}\\
Xavier~Costa-P\'erez,~\IEEEmembership{Senior Member,~IEEE}

\thanks{{\hspace{-0.3cm}Guillermo Encinas-Lago is with NEC Laboratories Europe, 69115 Heidelberg, Germany and Universit\'e Paris-Saclay, CNRS, CentraleSup\'elec, Laboratoire des Signaux et Syst\`emes, 91190 Gif-sur-Yvette, France; and i2Cat, 08034 Barcelona, Spain. \\
Vincenzo Sciancalepore is with NEC Laboratories Europe. \newline
Henk Wymeersch is with Chalmers University of Technology, Department of Electrical Engineering, 412 96 Gothenburg, Sweden. \newline
Marco Di Renzo is with Universit\'e Paris-Saclay, CNRS, CentraleSup\'elec, Laboratoire des Signaux et Syst\`emes, 91190 Gif-sur-Yvette, France. \newline
Xavier Costa-P\'erez is with i2cat, ICREA, and NEC Laboratories Europe. \newline
Email of the corresponding author: guillermo.encinas@i2cat.net.}}
}

This work has been submitted to the IEEE for possible publication.  Copyright may be transferred without notice, after which this version may no longer be accessible.

\newpage

\maketitle
\begin{abstract}
The advance towards 6G networks comes with the promise of unprecedented performance in sensing and communication capabilities. The feat of achieving those, while satisfying the ever-growing demands placed on wireless networks, 
promises revolutionary advancements in sensing and communication technologies. As 6G aims to cater to the growing demands of wireless network users, the implementation of intelligent and efficient solutions becomes essential. \change{In particular, \glspl{ris}, also known as Smart Surfaces}, are envisioned as a transformative technology for future 6G networks. 

\change{The performance of \glspl{ris} when used to augment existing devices is nevertheless largely affected by their precise location.} Suboptimal deployments are also costly to correct, negating their low-cost benefits. 
\change{This paper investigates the topic of optimal \glspl{ris} diffusion}, taking into account the improvement they provide both for the sensing and communication capabilities of the infrastructure while working with other antennas and sensors. 
We develop a combined metric that takes into account the properties and location of the individual devices to compute the performance of the entire infrastructure. We then use it as a foundation to build a reinforcement learning architecture that solves the \gls{ris} deployment problem.
Since our metric measures the surface where given localization thresholds are achieved and the communication coverage of the area of interest, the novel framework we provide is able to seamlessly balance sensing and communication, showing its performance gain against reference solutions, \change{where it achieves simultaneously almost the reference performance for communication and the reference performance for localization}.
\end{abstract}

\section{Introduction}
\label{sec:introduction}

Over the past decades, wireless communications and radar sensing have made significant advancements as prominent radio technologies. However, these technologies have developed independently with a very limited collaboration, despite their shared hardware architecture and signal processing principles~\cite{Liu22_JSAC}. Recently, the idea of \gls{isac} has emerged, garnering considerable attention in the research community~\cite{lui20_tcomm}. \gls{isac} aims to combine the functionalities of communication and sensing systems, leveraging scarce spectral resources to create cost-efficient, intelligent, and user-friendly wireless networks with unprecedented possibilities~\cite{zhang22_comst}.
Nevertheless, establishing a reliable \gls{isac} service poses serious challenges, \change{particularly in unfavorable electromagnetic (EM) propagation environments,} operating at millimeter-wave (mmWave) and terahertz (THz) frequency bands, which offer abundant spectral resources~\cite{gruber08_tap}. To overcome such limitations, researchers are currently exploring more sophisticated solutions: On the one hand, compensating for the path loss in high-frequency channels can be achieved through the use of massive \gls{mimo} systems~\cite{Abrardo21}, which provide high beamforming and spatial gains; on the other hand, \glspl{ris} have shown potential in mitigating the blockage effect among other problems~\cite{di2020smart}. Indeed, the groundbreaking \gls{ris} technology can programmatically alter the propagation properties of an incoming EM signal, thereby providing the means to control the surrounding propagation environment. Despite such promising techniques being extensively studied in separate fields of communications and sensing, their specific applications to \gls{isac} systems still remain relatively unexplored.

\begin{figure}[t]
        \center
      \includegraphics[width=\linewidth]{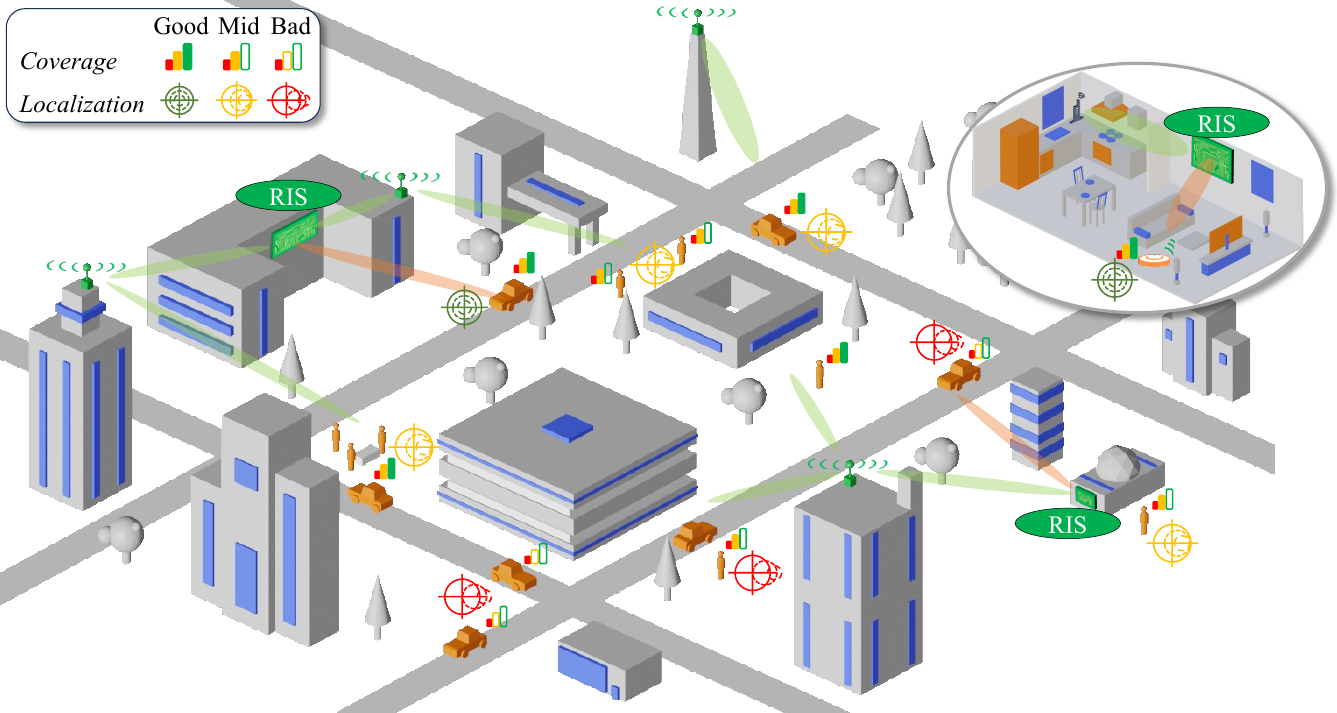} 
        \caption{\change{\gls{ris}-assisted \gls{isac} network simultaneously considering both indoor and outdoor scenarios.}}
        \label{fig:scenario}
\end{figure}

\glspl{ris}, proven to be low-complexity, low-cost, and passive devices, can play a major role in accomplishing the \gls{isac} revolution due to their sustainability and flexibility properties. However, installing and maintaining such innovative devices can  become an impelling task. \glspl{ris} must be properly placed and continuously configured to simultaneously $i)$ steer the communication beam towards the desired receiver and $ii)$ exploit the passive radar capabilities to collect real-time information from connected devices, which send timestamped pilot signals and infer the target position, as depicted in Fig.~\ref{fig:scenario}.

Nonetheless, during the initial roll-out phase, the placement of \glspl{ris} becomes of paramount importance: this has been recently studied in previous works that tackle the \gls{ris} deployment problem~\cite{albanese2022ris, encinas2023unlocking} from a theoretical perspective suggesting practical frameworks.  \change{In those practical frameworks, the performance gains provided by \glspl{ris} vary greatly with their placement. Additionally, the solutions to the deployment problem (i.e. the locations where \glspl{ris} are placed) can often be counterintuitive. These two properties (the importance and the counterintuitivity) of the deployment solutions, call for solid approaches to the issue. } 
While existing works attain to communications capabilities boost, they neglect other important benefits that networks obtain from incorporating \glspl{ris}. 
Conversely, in this work, we specifically aim to cover that gap, building a localization metric in seamless cooperation with a communications metric. We rely on such novel \glspl{kpi} while devising a practical \gls{ris} installation framework, as detailed in the following section.

\subsubsection*{Contributions}
In this paper, we propose \emph{\name{}}, a \gls{rl}-based method that, unlike existing solutions, maximizes \gls{ris} performance both in sensing and communication. In \name{} we explicitly consider environmental signal blockages and shadowing and treat \glspl{ris} and \glspl{bs} as sensors. 
To build it, we first transform the information from the sensors into consolidated spatial \glspl{pdf} of the user location estimations. Finally, 
computing the Fisher information contained in those \glspl{pdf}, 
we introduce meaningful metrics that serve as the heart of \name{}, jointly evaluating the localization and communication performance of deployments while allowing heterogeneous combinations of \gls{ris}, \glspl{bs}, and other sensors. 
 
Contributions are listed as follows: 
C$1$) a novel formulation of the \gls{ris}  deployment problem oriented to \gls{rl}, catering both sensing and communications performance, considering the simultaneous deployment of other devices (as \glspl{bs})
C$2$) mathematical models to describe the information obtainable from \glspl{ris}, predicting the device capabilities in an accurate and tractable manner, 
C$3$) a processing method to encode all the available information from sensors into a single Cartesian discrete \gls{pdf} of position estimations, 
C$4$) \change{a localization performance evaluation of the infrastructure based on Fisher information, able to drive the \gls{rl} algorithm, } 
C$5$) a joint sensing and communication metric for \gls{ris}, \gls{bs} and sensor deployments and 
C$6$) the performance evaluation of \name{}, leveraging on the previously listed contributions in a sample scenario.

\subsubsection*{Notation and conventions}
We denote vectorial quantities with boldface, and the modulus of an $N$-dimensional vector \change{$\textbf{v}$ as ${\Vert   \textbf{v}\Vert}$. We use} Cartesian coordinates when not otherwise noted. A notable exception is angular measurements of the direction of an impinging signal in an antenna, which we indicate using horizontal coordinates. In this context, we denote the \textit{elevation of arrival} by $\varphi_A$ and define it as the angle between the arrival direction and the reference plane $XY$. We denote the \textit{azimuth of arrival} by $\theta_A$ and define it as the signed angle between the reference direction $X$ and the projection of the arrival direction in the reference plane $XY$. Consistent with this, for simplicity we use the function $\text{atan}2(y,x)$, also in the vectorial form $\text{atan}2(\textbf{p}) \equiv \text{atan}2(y,x)$ with $\textbf{p} = (x,y,z)$, defined as the following:
\begin{equation}\label{eq:atan2} \nonumber
\text{atan}2(y, x) = \left\{ \begin{array}{cl}
 \arctan\left(\frac{y}{x}\right) &\text{if } x > 0, \\
 \frac{\pi}{2} - {\arctan}\bigl(\frac x y\bigr) &\text{if } y > 0, \\
 -\frac{\pi}{2} -{\arctan}\bigl(\frac x y\bigr) &\text{if } y < 0, \\
 \arctan\left(\frac y x\right) \pm \pi &\text{if } x < 0, \\
 \text{undefined} &\text{if } x = y = 0.
\end{array} \right.
\end{equation}

We use the Gaussian distribution, denoted as $\mathcal{N}(\mu, \sigma^2)$, where $\mu$ is the expected value or mean of the distribution and $\sigma^2$ is the variance.
We use $c$ for the speed of light in vacuum. We denote the natural logarithm of a quantity $q$ as $\ln(q)$ or simply $\ln q$, and the average of the same quantity as $\overline{q}$. For the expected value of a function $f(Z)$ of a random variable $Z$ we use the notation $\operatorname{\mathbb{E}}\left[f(Z)\right]$. The $i$-th eigenvalue of a matrix $M$ is denoted as $ev_i(M)$.

\section{Related Works}
The topic of \gls{isac} is currently receiving noticeable attention from academic literature. 
\change{We summarize here the publications that present the closest parallelism and relevance to the deployment of devices capable of simultaneous sensing and communications (an in particular, \glspl{ris}), in \gls{isac} systems. 

\subsection{\gls{isac} performance}}
The new trade-offs that arise when considering simultaneously sensing and communications aspects in the infrastructure are studied in works as~\cite{gan2024coverage} and~\cite{an2023fundamental}. These are taken into consideration when the theoretical limits of the capabilities of \gls{isac} networks are studied as in~\cite{gan2024coverage}, which can directly impact the deployment problem and other parallel questions, such as the power allocation problem studied in~\cite{an2023fundamental}.

\change{\subsection{Deployment problem in \gls{isac} systems}}
Very limited literature has addressed the optimal deployment problem considering both communication and sensing. For instance, \cite{albanese22_tmc} studied an optimal \gls{bs} deployment solution that jointly maximizes the overall system throughput and minimizes the \gls{crb}, thus improving the localization accuracy. We go one step beyond, looking at \gls{ris} as the emerging and groundbreaking technology. It has substantially attracted academic and industrial entities due to its versatility and agility to enhance existing network deployments. In the last few years, a plethora of research works have been published to tackle such novel technologies from different perspectives, including physical modeling, design and implementation, and optimization of \gls{ris} configuration and control. However, commercial exploitation involves a number of technical challenges due to the wide set of variables to take into account, such as operating frequency, size and orientation, or optimized \glspl{kpi}~\cite{Ma23_ojcoms}. 

\change{\subsection{\gls{ris} usage in \gls{isac} systems}}
The exploitation of the reconfigurable propagation environment that \glspl{ris} are able to build in the context of \gls{isac} systems is also well studied in the literature. For example, in~\cite{liu2023integrated} the authors describe the topic and the capabilities \glspl{ris} offer for \gls{isac}, as well as documenting the state of the art on the techniques studying how to optimally incorporate these devices along existing machines.
In~\cite{chen21_twc}, the authors proposed a new \gls{ris}-aided edge caching system by formulating a network cost minimization problem to jointly optimize content placement at cache units, active beamforming at \gls{bs}, and \gls{ris} configuration. This is performed by exploiting an alternating optimization algorithm to jointly tackle the \gls{bs} beamforming and \gls{ris} configuration.

\change{\subsection{\gls{ris} deployment problem}}
In \cite{zeng21_lcomm}, the authors analyzed the coverage of a downlink \gls{ris}-assisted network. The \gls{ris} placement optimization problem was formulated to maximize the cell coverage by optimizing the \gls{ris} orientation and horizontal distance with the \gls{bs}.
In~\cite{ntontin21_ojcom}, an optimal \gls{ris} placement for highly directional mmWave links is presented. The authors highlighted the relationship between transmission beam footprint at the \gls{ris} plane and its corresponding size. The authors of~\cite{hou22_ictc} \change{focused on the mmWave communications} in indoor scenarios. They formulated a \gls{ris} placement optimization to maximize the coverage area with a supervised learning approach outperforming the conventional schemes attained by the decision tree algorithm and the randomly deployed \gls{ris}. Finally, \cite{ntontin21_eucap} evaluated the end-to-end \gls{snr} expression of the transmitter-\gls{ris}-receiver links to acquire important insights about the \gls{ris} position on the overall system performance.

Moving towards higher frequencies, such as D-band, in~\cite{stratidakis22_tap} the optimal placement of the \gls{ris} is studied with respect to both position and orientation. In particular, the proposed analysis is based on an analytical model that treats the \gls{ris} as a continuous surface of finite size that steers the incident beam toward a prescribed direction. 

\hspace{0.1mm}

\change{Nevertheless, none of these approaches looked into the optimal \gls{ris} installation and configuration problem to pursue both localization and communication performance maximization at the same time, leading us to work within that niche in the present work.
}

\section{System Model}
\label{sec:system_model}

\begin{figure}[t]
        \center
        \includegraphics[width=\linewidth, trim = {0cm 9cm 5cm 0cm}]{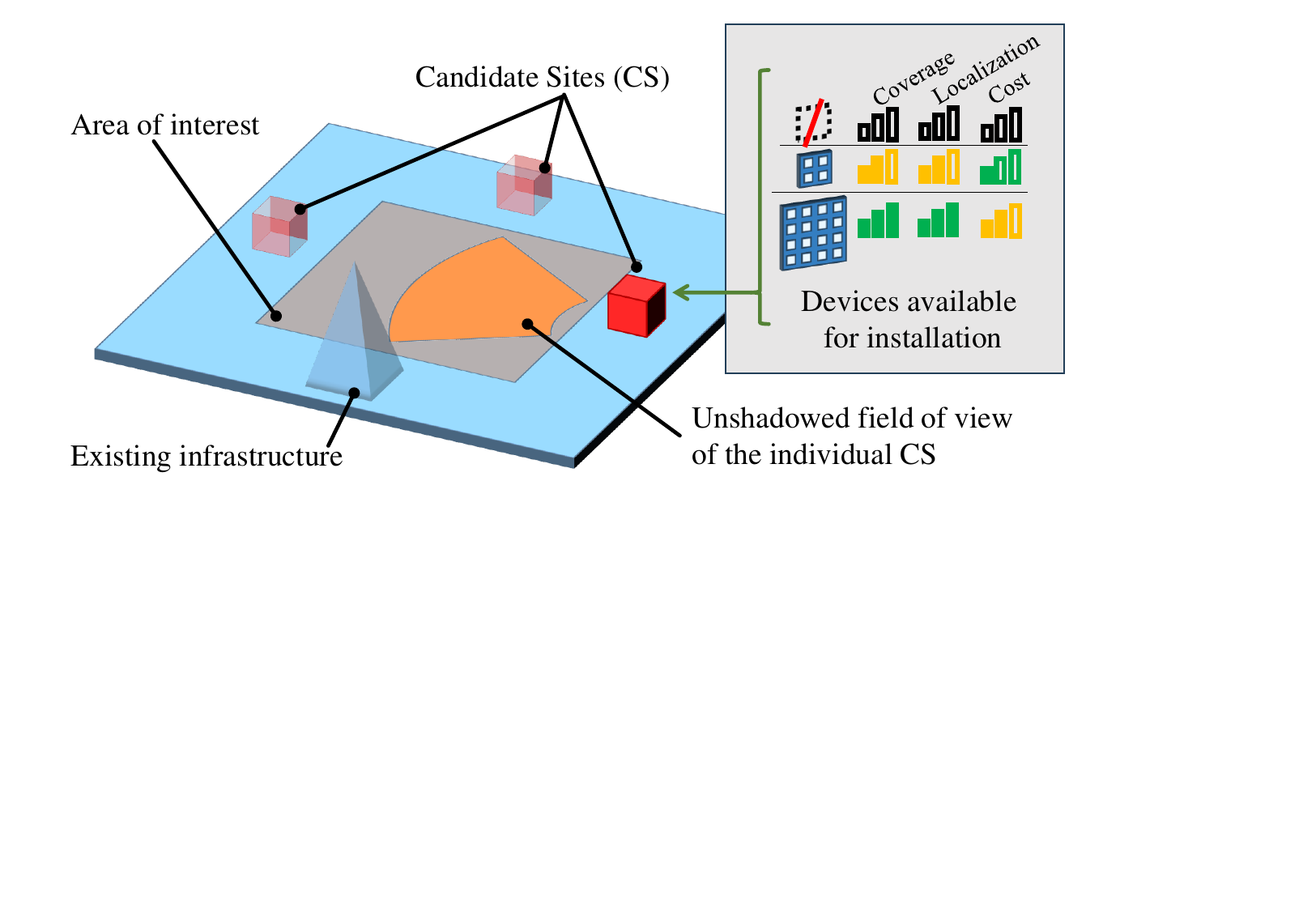}
        \caption{Schematic view of the elements of the system model.}
        \label{fig:system_model}
\end{figure}

To define the initial problem and develop our proposed solution, we examine a traditional propagation environment where transmitters and receivers can establish connections via both direct links (i.e., \gls{los}) and reflected links (i.e., utilizing the conventional properties of the \gls{ris}). We consider a confined scenario, where we intend to deploy devices to improve coverage and localization services while taking into account the existing infrastructure. This is depicted in Fig.~\ref{fig:system_model}. 

\change{In this section we explain our \textit{scenario}, detailing the model where we base our work. We explain how we model the devices in the infrastructure able to provide information (\textit{sensing nodes}), and how we encode that information. We consider two particular examples, inspired in \glspl{ris}: \textit{\gls{toa} sensing} and \textit{\gls{aoa} sensing}. We explain also the way we model \textit{shadowing} in our scenario. }

{\subsection{Scenario}} \label{'sec:Scenario'} We study an area where multiple \glspl{ue} are placed. In this area we want to deliver boosted-performance communication and sensing services by means of a wireless network, as depicted in Fig.~\ref{fig:scenario}.
To achieve this we consider that we can either install \glspl{ris} into the environment (in order to enhance the existing wireless infrastructure, e.g., access points or \glspl{bs}) or by planning a joint \glspl{bs} and \glspl{ris} deployment. Our intention is to exploit the controllable (passive) beamforming at the \gls{ris} devices that, in turn, become $i$) effective passive sensors able to accurately detect the position of a target user (please refer to Theorem~\ref{theorem:ris} in Section~\ref{sec:math}) and $ii$) passive signal repeaters to significantly enhance the overall communications while ensuring the area receives a sufficient level of coverage at all points. 
The technical challenge relies on finding the optimal overall network deployment that maximizes both communication and sensing performance. 

In our study, we consider devices that are deployed with varying characteristics: this includes differences in the number and arrangement of antennas, the accuracy of their sensing capabilities, and, in the case of \glspl{bs}, the transmitted power.
To better consider the characteristics of the physical environment of our study, we take into account the presence of shadowing between each potential deployment site, namely \gls{cs} and the area where \glspl{ue} are distributed. We represent this shadowing information as a collection of blockage areas for each potential \gls{cs}, and we encode the environment propagation conditions using such blockage areas. This is illustrated in Fig.~\ref{fig:system_model}, where the shadowed area of the highlighted \gls{cs} is colored gray and the unshadowed (usable) area is colored orange. It is also further explained and justified in Sec.~\ref{sec:system_model} subsection \textit{E. Shadowing and connection models}. 
\change{Additionally, we can incorporate to our method information about the expected statistical distribution of \gls{ue} locations in the scenario. In our implementation we consider the worst case for the localization accuracy, a uniform distribution of users, representing the lack of prior information.}
We exploit the information on shadowing areas to create a model of the propagation environment in which our solution will operate. This model helps us simulate how each solution will perform in practice. On the other hand, the statistical distribution of \gls{ue} locations is utilized in the evaluation metrics.\footnote{This can be used, for example, by applying weights to the metric evaluations based on the density of \glspl{ue} in different spatial regions and collecting corresponding results, as shown in Section~\ref{sec:results}.}
By using the statistical distribution of \glspl{ue}, we can effectively exclude the evaluation of areas that are inaccessible or not relevant for our study: this simplifies the evaluation process and allows us to focus on the regions where the solutions are viable and applicable with a relevant output.

\begin{figure}[t]
        \center
        \includegraphics[width=\linewidth, trim= {3.5cm 9.5cm 0cm 0cm}]{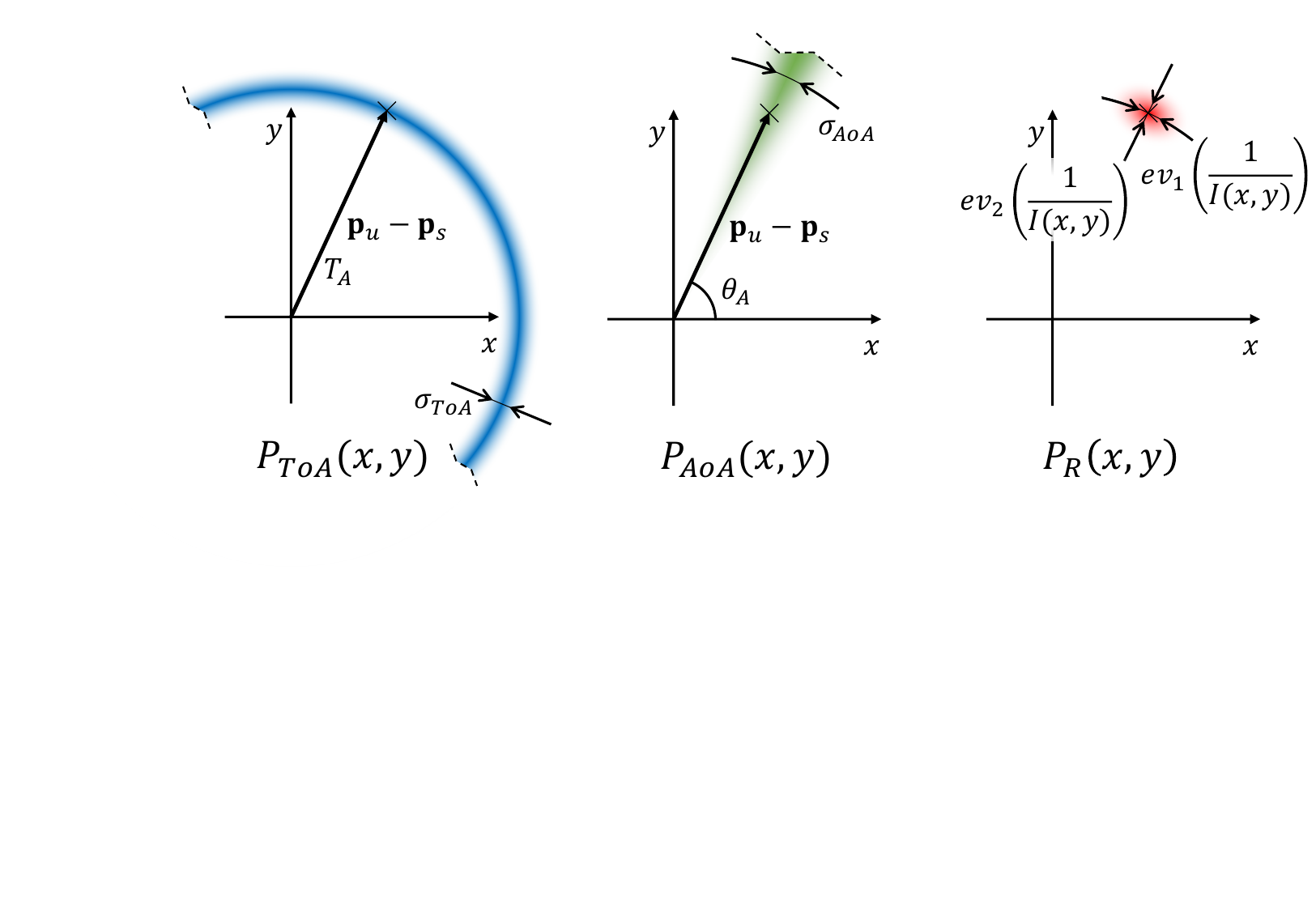} 
        \caption{Time of arrival (ToA) (left) and angle of arrival (AoA) (center) measurements diagrams in 2D, with their combination (right).}
        \label{fig:pdf_diagrams}
\end{figure}

{\subsection{Sensing nodes}} \label{'sec:Sensing nodes'} Fundamentally, the construction of \name{} allows for the acceptance of any sensing information (i.e., the output from any sensor) that can be projected into a Cartesian coordinate space. This would include, for example, pyroelectric infrared presence sensors within a room~\cite{lau2017sensor}, mmWave radars~\cite{wei2022mmwave}, channel information fingerprinting~\cite{wang2015deepfi}, or even secondary information sources, such as location information inferred from sound noise levels, energy consumption, etc. For conciseness and clarity, in this work we focus on the deployment of \glspl{ris}---one of the most relevant technologies for sensing and communications~\cite{wen2024shaping}---and \glspl{bs}. We model and use the information readily obtainable from them as the basis for \name{}. We assume that each device (i.e., \glspl{bs} and \glspl{ris}) can provide sensing information in the form of either \gls{toa}, \gls{aoa}, or both, as shown in the left and center parts of Fig.~\ref{fig:pdf_diagrams}. On the right part of this figure we also introduce elements properly covered later in the paper, as the combination of the information from all the individual sensors (detailed in Sec.~\ref{sec:math}) and the extraction of the accuracy achievable from this (detailed in Sec.~\ref{sec:solution}. We infer the individual measurement accuracy of each sensor from the intrinsic characteristics (e.g., the number and distribution of elements of a \gls{ris}), the operating frequency, the measurement conditions, e.g., the \gls{snr} at the sensor doing the measurement, etc.~\cite{zhang2021performance} With the accuracy of the sensors, we model the distribution of the measurements expected from each sensor as a Gaussian distribution, where the average $\mu$ is the real value of the measured magnitude and $\sigma^2$ is the variance of the statistical distribution of the measurements obtained from the sensor. 
\gls{toa} and \gls{aoa} sensors have different models for the dependence of their accuracy with the \gls{snr}. In the following, we describe how \gls{toa} and \gls{aoa} sensing procedures are performed.

\vspace{2mm}
{\subsection{\gls{toa} sensing procedure}}  \label{'sec:ToA sensing'} We model the \glspl{bs} placed in the scenario as nodes with \gls{toa} sensing capabilities. If we consider a sensor in the location $\textbf{p}_s = (x_s,y_s,z_s)$ \change{with an internal clock synchronized with the reference time $t$,} and a \gls{ue} located at $\textbf{p}_{u} = (x_{u},y_{u},z_{u})$ with an internal clock $t_{u}$ presenting an unknown constant deviation from the reference time $t$ so $t-t_{u} = \Delta_{t_{u}} \neq 0$, a \gls{toa} measurement translates to a spherical cone: 
\begin{equation}\label{eq:hypercone}
(T_A-\Delta_{t_{u}})^2 \cdot c^2 = (x_{u}-x_s)^2 + (y_{u}-y_s)^2 + (z_{u}-z_s)^2,
\end{equation}
\change{where the measurement $T_A$ is the difference between the absolute time of arrival of the sounding signal (in the time reference system) and the \gls{ue} time of emission of said signal (in the \gls{ue} own time reference).}
\change{Eq.~\eqref{eq:hypercone} presents a simple concept: the right term represents the distance between the \gls{ue} and the sensor that performs the \gls{toa} measurement, squared. The left term represents the time of flight of the measurement signal multiplied by the speed of the signal, $c$, both squared.}
This turns into the following simpler form
\begin{equation}\label{eq:simple_circle}
{T_A}^2 \cdot c^2 = (x_{u}-x_s)^2 + (y_{u}-y_s)^2,
\end{equation}
when considering a scenario wherein we are restricted to a known, constant height and where the internal clock deviation of the \gls{ue} is assumed to be negligible~\footnote{Neither of these assumptions is required formally nor computationally for our method: we exploit them to ease the exposition both in terms of formulas and graphical depictions.}. 

For the \gls{toa} measurement accuracy we take the value of the range accuracy of a secondary radar using a bandwidth $B$~\cite{Curry_2005}:
\begin{equation}\label{eq:ToA_sigma}
\sigma_{ToA} = \frac{c}{2\cdot B \sqrt{2\cdot \text{SNR}}} .
\end{equation}

We model \glspl{ris} as passive devices without sensing or computational capabilities (\!\!\cite{Ma23_ojcoms}), and hence they do not offer any direct feedback. Nevertheless, we can obtain implicit information from their operation: the alterations in observable environmental variables---e.g. the \gls{csi}---following controlled (or known) changes in the \gls{ris} configuration can be exploited to gain knowledge about the environment or, specifically in our case, the position of the \gls{ue}~\cite{di2020smart}. We assume the \gls{bs} is able to discriminate the received signals in order to distinguish those coming from the \gls{ue} and the ones programmatically reflected by the \gls{ris}, thereby extending \gls{bs} \gls{toa} sensing capabilities to those devices. \change{A \gls{bs} could distinguish those, for example, by the \glspl{toa} themselves, the rough \gls{aoa} of the incoming signals, or other factors~\cite{1054120}}. \glspl{ris} then are considered as anchors, with a decrease in the accuracy due to the additional reflection and the increased path length. We then use the same value for the time resolution and, by extension, the same $\sigma^2_{min,ToA}$ as for the \glspl{bs}, but we do update the \gls{snr} of the measurement to reflect the accuracy degradation. 

{\subsection{\gls{aoa} sensing procedure}} \label{'sec:AoA sensing'} We model \gls{ris} devices with the capability to quickly re-configure, able to follow and focus on the \gls{ue} they are currently serving. This operating regime has the benefit of offering precise angular information~\cite{he2020adaptive}. If we consider a sensor in the location 
$\textbf{p}_s = (x_s,y_s,z_s)$ and a \gls{ue} located at $\textbf{p}_{u} = (x_{u},y_{u},z_{u})$, an \gls{aoa} measurement in horizontal coordinates translates into the following equations: 
\begin{align}
    \theta_A = \text{atan}2(y_{u}-y_s, x_{u}-x_s), \label{eq:3dline_theta}\\ 
    \varphi_A = \text{arcsin}(\frac{z}{\sqrt{x^2+y^2+z^2}}), \label{eq:3dline_phi}
\end{align}
where the values $\varphi_A$ and $\theta_A$ are the central values of angular measurements of the \gls{aoa} in the local horizontal coordinate system of the sensor. These equations simply come from the definition of horizontal coordinates, where the elevation $\varphi_A$ is measured from the horizontal plane $XY$ or $Z=0$, and the azimuth angle $\theta_A$ employs the $\text{atan}2(y,x)$ function, which is defined specifically for this purpose as described in Section~\ref{sec:introduction}. Since the statistical distribution of measurements is assumed unbiased, the central values of the measurement distributions are the true values.

We model the accuracy of the sensor in any given measurement as equal to its base accuracy, with a degradation factor dependent on the \gls{snr}~\cite{keykhosravi2022ris}.
If the sensor can achieve in optimal conditions a minimum value $\sigma_{min}^2$, we model the accuracy dependence with the \gls{snr} by dividing that value with a sigmoid function ranging between the value $0$ at the minimum \gls{snr} (with a sensitivity threshold $\text{SNR}_{min}$), and $1$ at the saturation \gls{snr} beyond which there are no noticeable improvements in signal quality, noted as $\text{SNR}_{max}$. Having chosen the logistic function, the sensor variance at the measurement $\sigma^2_{AoA}$ yields 
\begin{align}
\sigma^2_{AoA} & = \sigma_{min}^2 \cdot \frac{1+e^{k_{\text{SNR}}(\text{SNR}_m-\text{SNR}_c)}}{e^{k_{\text{SNR}}(\text{SNR}_m-\text{SNR}_c)}},\label{eq:sensor_var}\\
\text{SNR}_c &= \frac{\text{SNR}_{max} + \text{SNR}_{min}}{2},\label{eq:SNR_c}\\
k_{\text{SNR}} &= \frac{2 \cdot \ln(9) }{(\text{SNR}_{max} - \text{SNR}_{min})},\label{eq:k_SNR}
\end{align}
where $\text{SNR}_m$ represents the \gls{snr} obtained by the sensor at the measurement, and we choose the factors $k_{\text{SNR}}$ and $\text{SNR}_c$ to ensure the intended behavior of the model. What we intend with this model for $\sigma^2_{AoA}$ is to encode the dependence of the sensor reliability with \gls{snr}. In Eq.~\eqref{eq:sensor_var} for high values of $\text{SNR}_m \geq \text{SNR}_{max}$ the sensor accuracy tends to the optimal value $\sigma^2_{AoA} \approx \sigma_{min}^2$, and for low values when $\text{SNR}_m < \text{SNR}_{min}$ we the accuracy falls as $\sigma^2_{AoA} \gg \sigma_{min}^2$, with a smooth transition in between. This smooth transition is governed by the factors $\text{SNR}_c$ and $k_{\text{SNR}}$: with the values chosen in Eq.~\eqref{eq:SNR_c} and Eq.~\eqref{eq:k_SNR} we have $\sigma^2_{AoA} = \sigma_{min}^{2} / 0.9$ when $\text{SNR}_m = \text{SNR}_{max}$, representing near-optimal operation and $\sigma^2_{AoA} = \sigma_{min}^{2} / 0.1$ when $\text{SNR}_m = \text{SNR}_{min}$, representing the absence of usable information in that circumstance.

We take the minimum variance $\sigma_{min}^2$ of the measurements as the sensor resolution. The resolution in our model depends on the shape of the beam-pattern produced by the \gls{ris}, particularly the beam-width. Hence, the defining characteristic of the resolution of the \gls{ris} sensor will ultimately be the number and distribution of antenna elements on that device. To maintain a consistent form with the sensor Gaussian model detailed before, we model this beam-width as half the angular span between the two closest points to the beam centre that present a gain equal to $G_{BW} = G_{max} \cdot e^{-\frac{1}{2}}$, where $G_{max}$ is the gain in the centre of the beam. As for \glspl{bs}, we distinguish between fixed antennas, without a precoder or any other means to produce a configurable directional beam, and \gls{mimo} \glspl{bs}, composed of an \glspl{ura}, \glspl{uca} or other similar means able to focus on the individual users at will, and hence producing angular information similar to that produced by \glspl{ris}.

{\subsection{Shadowing and connection models}} \label{'sec:Shadowing'} 
\change{We consider a known environment, were we can evaluate whether a pair of points are linked through a \gls{los} or not.
We take into account each \gls{cs} individually, and check the availability of \gls{los} with the all the points included in the area where \glspl{ue} are expected to be distributed.}
The areas where this \gls{los} condition is not fulfilled are noted as blockage areas and they are listed for each \gls{cs}. 
Considering the propagation characteristics of the radio frequency bands where the \gls{ris} is an effective solution, \gls{nlos} links caused by reflections in the environment are excluded, as their power contribution is minimal under these conditions.
This dichotomous behaviour is not only theoretically predicted but it has been also widely observed in simulations and field measurements~\cite{xing2021millimeter, sulyman2014radio}, providing further justification for our binary shadowing model. Having taken into account blockage with this mechanism, we can model remaining \gls{los} connections with a simple, low-computational-complexity form of the Friis transmission equation:
\begin{equation}\label{eq:friis}
P_r = P_t \cdot G_t \cdot G_r \cdot \frac{\lambda^2}{(4 \pi d)^\gamma},
\end{equation}
where $P_t$ and $P_r$ are the transmitted and the received power, $G_t$ and $G_r$ are the transmitting and receiving gains, $\lambda$ is the wavelength used for the transmission, $\gamma$ is the path loss exponent, and $d$ is the distance between the transmitting and the receiving antennas, all in metric units. While this simple model allows us to focus our work on the localization aspect, it is worth mention that the \gls{rl} architecture we propose can be swiftly adapted to employ models with arbitrarily high detail, as long as they can provide a map of the chosen parameter to evaluate the communication quality (in our case the \gls{snr}), given a fixed infrastructure. For example, if a detailed and realistic simulator as the one described in ~\cite{sihlbom2022reconfigurable} is available, our proposal can integrate its output to obtain much more refined solutions on the communications side, without any further adaptation or change needed on our proposal. \change{These improvements on the simulation of individual scenarios could be expensive while not offering enough insight about the general case, and thus might be unfit for general research. Nevertheless the higher reliability of the data obtained from more detailed simulations can be the basis of a commercial exploitation of both the present work and ~\cite{sihlbom2022reconfigurable}. }

\section{Statistical positioning quality}
\label{sec:math}
In this section we detail the fundamentals of our solution from a mathematical perspective. In particular, we consolidate our proposal premises and provide a solid guideline that allowed us to design in the next section our novel \gls{ris}-based \gls{isac} framework.  

{\subsection{Measurement projections}} Given the heterogeneity of available measurements, we need to tailor all such information onto the same space. Hence, we map each one-dimensional measurement with its variance into the full-dimensional space of states for the \gls{ue}, which includes $3$ spatial dimensions (e.g., for the \gls{ue} position, $x_{u}, y_{u}, z_{u}$) and one time dimension (e.g., for the \gls{ue} internal clock deviation $\Delta_{t_{u}}$) in the original form, or only $2$ spatial dimensions $(x_{u}, y_{u})$ in the simplified form. 

The result of the mapping is one analytical \gls{pdf} for each measurement, assuming a Gaussian distribution of errors with the given standard deviation, individual to each measurement and dependent on the characteristics of the sensing device and measurement conditions. Since we assume the measurements to be normally distributed, all the information they carry is encoded in their central value $\mu$ and variance $\sigma^2$. Since the mapping between a pair of values $(\mu, \sigma^2)$ of any given measurement and their corresponding \gls{pdf} is surjective we consider it a projection. With these \glspl{pdf} we express the measurement \textit{relative likelihood} of producing a value, given the full state definition. 
The \gls{pdf} $P(\textbf{s}|m)$ spanning over all the space of states $\textbf{s}$ conditioned to a measurement real value $m_r$ can be expressed as the following
\begin{equation}
    P(\textbf{s}|m_r) \propto P_s( \widehat{m}(\textbf{s})|m_r),
\end{equation}
where $P_s(\widehat{m}|m_r)$ is the relative likelihood of obtaining a measurement $\widehat{m}$ from a sensor given the real value $m_r$, and $\widehat{m}(\textbf{s})$ is the measurement associated with state $\textbf{s}$.
 
While both the mathematical derivations and the computational counterparts can be performed in the original $\mathbb{R}^4$ space, for simplicity of the exposition we show here the $2$-dimensional \glspl{pdf} where we assume the \glspl{ue} are located in $z_{u}=z_c$ plane while discarding the \gls{ue} internal clock deviation (i.e., $\Delta_{t_{u}} = 0$) and the \gls{aoa} elevation measurements $\varphi_A$. As a measurement value can be produced by infinite points in the state space, we need more information to create the projection from $\mathbb{R}$ to $\mathbb{R}^n$. We assume that the state space is equiprobable\,\footnote{This assumption makes our analysis more tractable leading to an \gls{ue} located in any position and have any internal clock deviation within the considered space with the same probability.}, and hence all possible states producing the same measurement are equiprobable too.

{\subsection{Location probability distribution}} Let us consider a sensor in the location $\textbf{p}_s = (x_s,y_s,z_s)$ and a \gls{ue} located at $\textbf{p}_{u} = (x_{u},y_{u},z_{u})$. If we derive the \gls{pdf} encoding the spatial information from an \gls{aoa} measurement $\widehat{\theta}$ using Eq.~\eqref{eq:3dline_theta} and the Gaussian distribution, it yields

\begin{align}
P_{AoA}(x,y) & = C_{N,AoA} \cdot e^{\frac{-1}{2} \cdot \frac{D^2_{AoA}(x,y)}{\sigma^2_{AoA}}},\label{eq:AoA_PDF} \\ 
D_{AoA}(x,y)&  = \widehat{\theta}-\theta_r, \label{eq:AoA_D}\\ 
\widehat{\theta} & = \text{atan}2(y-y_s, x-x_s),\\
\theta_r & = \text{atan}2(y_{u}-y_s, x_{u}-x_s),\label{eq:AoA_Tr}\\
C_{N,AoA} & = \left(\int_{\mathbb{R}}\int_{\mathbb{R}} e^{\frac{-1}{2} \cdot \frac{D^2_{AoA}(x,y)}{\sigma^2_{ToA}}} \ dx \ dy \right)^{-1}, \label{eq:AoA_C}
\end{align}
as depicted in Fig.~\ref{fig:pdf_diagrams} on the left-side hand in blue color. The intuition behind equations Eq.~\eqref{eq:AoA_PDF}-\eqref{eq:AoA_C} is simple: for any given angular measurement $\widehat{\theta}$ of the user position, the user has a probability density $P_{AoA}(x,y)$ of being located in the point $(x,y)$. To build this we simply consider that our angular measurements $\widehat{\theta}$ will have a normal distribution. Hence, as written before, we take the Gaussian distribution function, center it at $\theta_r$ following Eq.~\eqref{eq:3dline_theta}, and spread it accordingly to the expected sensor accuracy $\sigma^2_{AoA}$ from Eq.~\eqref{eq:sensor_var}. That leaves us with  $\mathcal{N}(\theta_r, \sigma^2_{AoA})$, as seen in Eq.~\eqref{eq:AoA_D} with the only additional factor of the normalization constant $C_{N,AoA}$. Thanks to this factor, defined in Eq.~\eqref{eq:AoA_C}, we normalize the total probability of $\int_{\mathbb{R}}\int_{\mathbb{R}} P_{AoA}(x,y) \ dx \ dy = 1$. Similarly to this, the \gls{pdf} encoding the spatial information from an \gls{toa} measurement $\widehat{T_A}$ using Eq.~\eqref{eq:simple_circle} and the Gaussian distribution can be written as the following

\begin{align} 
P_{ToA}(x,y) & = C_{N,ToA} \cdot e^{\frac{-1}{2} \cdot \frac{D^2_{ToA}(x,y)}{\sigma^2_{ToA}}},  \label{eq:ToA_PDF} \\
D_{ToA}(x,y) & = (\widehat{T_A} - T_{A,r}),\label{eq:ToA_D} \\  
\widehat{T_A} & = \sqrt{(y-y_s)^2 + (x-x_s)^2} / c\\
T_{A,r} & = \sqrt{(y_{u}-y_s)^2 + (x_{u}-x_s)^2} / c \\
C_{N,ToA} & = \left(\int_{\mathbb{R}}\int_{\mathbb{R}} e^{\frac{-1}{2} \cdot \frac{D^2_{ToA}(x,y)}{\sigma^2_{ToA}}} \ dx \ dy \right)^{-1}, \label{eq:ToA_C}
\end{align}
depicted (unnormalised) in Fig.~\ref{fig:pdf_diagrams} in the center in green color. 
As per the previous group, Eqs.~\eqref{eq:ToA_PDF}-\eqref{eq:ToA_C} allow us to compute the probability density $P_{ToA}(x,y)$ of the user being in a position $(x,y)$, given this time a \gls{toa} measurement $\widehat{T_A}$. The construction of these equations to model the \gls{pdf} for the \gls{toa} measurements is analogous to the previous case in Eqs.~\eqref{eq:AoA_PDF}-\eqref{eq:AoA_C}, in this case obtaining $\widehat{T_A}$ from Eq.~\eqref{eq:simple_circle} and $\sigma^2_{ToA}$from Eq.~\eqref{eq:ToA_sigma}. We have again a normalised total probability of $\int_{\mathbb{R}}\int_{\mathbb{R}} P_{ToA}(x,y) \ dx \ dy = 1$ thanks to the factor $C_{N,ToA}$.
As we consider the distributions of the internal errors of the sensors to be independent between devices, we can combine the available information by multiplying the \glspl{pdf} as the following

\begin{equation} \label{eq:prod_PDF}
P_R(x,y) = C_{N,R} \cdot \prod_i P_i(x,y),
\end{equation}
and re-normalising the result using the variable $C_{N,R}$ to keep
\begin{equation} \label{eq:norm_PDF}
\int_{\mathbb{R}}\int_{\mathbb{R}} \prod_i P_i(x,y) \ dx \  dy = 1,
\end{equation}
which we depict for one \gls{toa} \gls{pdf} and one \gls{aoa} \gls{pdf} in Fig.~\ref{fig:pdf_diagrams} on the right-hand side in red color. It is worth noting that, unlike the \gls{toa} \gls{pdf} written in Eq.~\eqref{eq:ToA_PDF}, the \gls{aoa} \gls{pdf} written in Eq.~\eqref{eq:AoA_PDF} is equal to zero. This is because as per Eq.~\eqref{eq:AoA_PDF} the normalization constant collapses to $C_{N,ToA} = 0$ since the double integral diverges, as it can be seen in the center of Fig.~\ref{fig:pdf_diagrams}, where the (infinitely spanning) \gls{pdf} is cut with a dashed line. Yet, when multiplied by other distributions as in Eq.~\eqref{eq:prod_PDF}, either angular as Eq.~\eqref{eq:AoA_PDF} itself (but with different values for $\textbf{p}_s$ or $\textbf{p}_{u}$), or radial as Eq.~\eqref{eq:ToA_PDF}, and then renormalised as Eq.~\eqref{eq:norm_PDF}, its information is preserved and traduced to the final result. With this in mind, we can formulate the following theorem.

\begin{theorem}
For a given pair of sensor accuracies in \gls{aoa} and \gls{toa} measurements $(\sigma_{ToA}, \sigma_{AoA})$, and with the objective of maximizing localization accuracy, \gls{ris} behave as an optimally placed pair of sensors.
\label{theorem:ris}
\end{theorem}

\begin{sketch}
Consider a PDF $f$, which encodes a set of measurements $\mathcal{\varepsilon}$ attaining a localization precision of $P$.
Let $\lambda_i$ be the eigenvalues of $\mathcal{I}(\textbf{q})$, the Fisher information matrix of $f$, which provide an upper bound for $P$, i.e. $\max_{\varepsilon}\{P\}\leq \lambda_i$.
Following Eq.~\eqref{eq:AoA_PDF} and Eq.~\eqref{eq:ToA_PDF}, we define two different \glspl{pdf}, $f_{ToA}:\mathbb{R}^4\rightarrow \mathbb{R}$ and $f_{AoA}:\mathbb{R}^3\rightarrow \mathbb{R}$, describing \gls{toa} and \gls{aoa} measurements respectively. We denote by $f_R(x,y,z,t)$ the properly normalized PDF obtained by the product of $f_{ToA}$ and $f_{AoA}$.
Lastly, assume that both original measurements are obtained from two sensors whose positions, given by $\vec{p}_{s,ToA}, \vec{p}_{s,AoA}$, are pair-wise independent. Then, $\max_{\vec{p}_{s,ToA}, \vec{p}_{s,AoA}} \lambda_i = \lambda_{\vec{p}_{s,ToA} = \vec{p}_{s,AoA}}$, since the eigenvectors associated to the largest eigenvalues of the Fisher information matrices of $f_{ToA}$ and $f_{AoA}$ need to be spatially orthogonal. \hfill \IEEEQEDopen

\end{sketch} 

\begin{figure}[t]
        \center
      \includegraphics[width=\linewidth, trim = {0cm 10cm 0cm 0cm}]{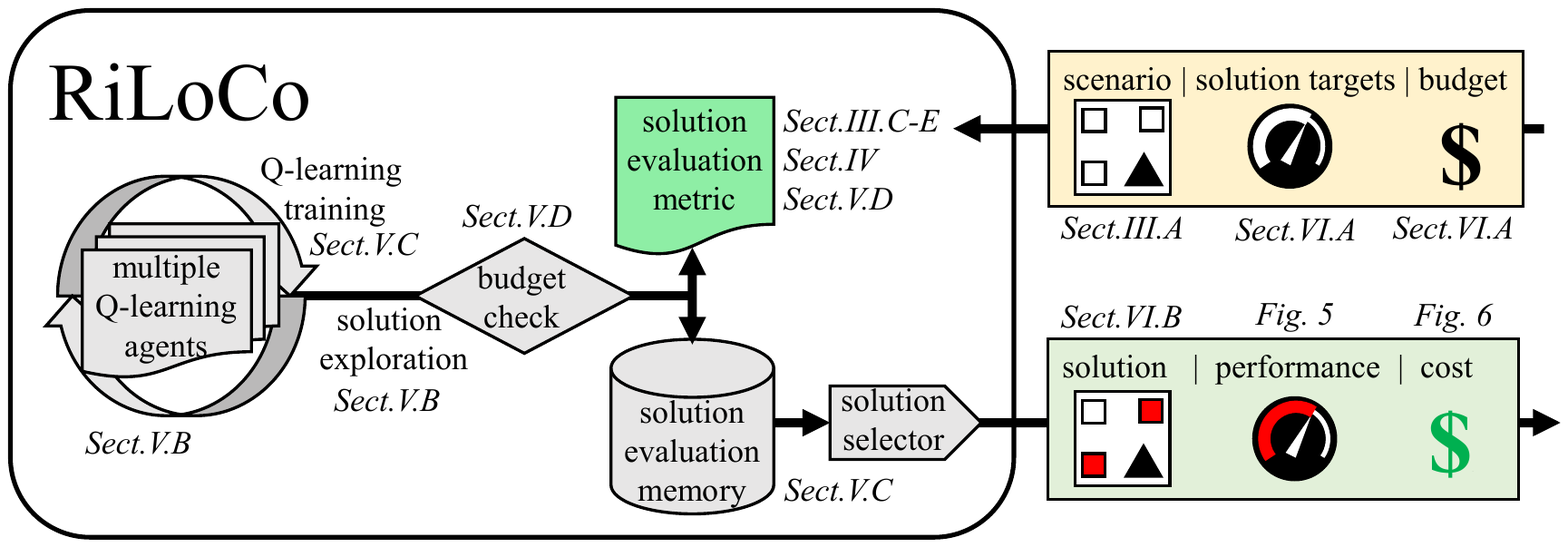}  
        \caption{Overview of \name{} building blocks.}
        \label{fig:blocks}
\end{figure}
We provide an extensive description of our framework by shedding the light on the main novel building blocks and their corresponding interactions within a classical propagation environment. 

\subsection{Problem definition}
We consider an area $\mathcal{A}_{u}$ with an extension $A_{u}$ where we aim at providing communication and localization services to a homogeneous distribution of \glspl{ue}. We have $N$ pre-defined \gls{cs} available for the installation of \glspl{ris} or \glspl{bs}. We define the deployment vector as $\bold{d}=\{d_i\}^{N}$, with $d_i$ being the device installed the $i$-th \gls{cs}. We have a maximum budget constraint $\beta$ such that the overall installation cost is within such a budget and a function $\kappa(\bold{d})$ that economically evaluates any deployment solution. \change{This budget can represent economical means or energy consumption, in order to optimize the energy efficiency of the infrastructure.} We have a set of devices $\mathcal{V} = \left\{0,1, ..., V\right\}^{N}$, each with a set of characteristics, available to be installed in each \gls{cs}, and including an entry to represent the absence of any installed device. We use two initial metrics in our proposal. The localization metric is denoted as $L(\bold{d},x_{u},y_{u})$, and the communication performance metric is denoted as $C(\bold{d},x_{u},y_{u})$. We detail both metrics later in this section, subsection \textit{D. Performance metrics}, with their individual descriptions highlighted in bold as \textbf{Communication} and \textbf{Localization}.  We define the joint metric as 
\begin{equation} \label{eq:joint_metric}
M(\bold{d},x_{u},y_{u}) = \alpha \ L(\bold{d},x_{u},y_{u}) + (1-\alpha) \ C(\bold{d},x_{u},y_{u}),
\end{equation}
where $\alpha$ marks the balance between localization and communications. With $\alpha = 1$ the joint metric attends only to localization and $\alpha = 0$ attends only to communications. Now, we can formulate our optimization problem as the following 

\begin{problem}[Joint Sensing and Communications Optimization]\label{problem:joint_optimization}
\begin{subequations}\label{eq:joint_objective_collective}
{\begin{align}
   & \max_{\bold{d}} && \frac{1}{A_{u}} \ \underset{\mathcal{A}_{u}}{{\iint}} M(\bold{d},x_{u},y_{u}) \, dx \ dy \label{eq:joint_objective}\\
   & \textup{subject to} && \kappa(\bold{d}) \leq \beta,    \label{eq:constraint_a_leq_beta}\\
   & && \bold{d} \in \left\{0,1, ..., V\right\}^{N}. \label{eq:constraint_sum_a_leq_1}
\end{align}}
\end{subequations}
\end{problem}

Note that, in our formulation the budget observed in the constrain Eq.~\eqref{eq:constraint_a_leq_beta} can be also easily incorporated in our metric Eq.~\eqref{eq:joint_metric} as a divisor, allowing to not only produce solutions optimizing the performance per watt, or abiding a capital expenditure ceiling, but also to search for the best \gls{roi}, among other energy-efficiency and economically-guided solutions, as later shown and detailed in Section~\ref{sec:results}.

\subsection{\name{} Architecture}

{\bf Overview.} The existing formulations of the deployment problem are, even without taking \glspl{ris} into consideration, NP-Hard~\cite{nguyen2018mobile}. Hence, finding solutions that cover novel situations and paradigms in a practical, feasible way is an open problem. Nevertheless, in this paper the solving algorithm itself occupies a secondary position, as the core of the presented work is the metric that we build in this section, in the subsection \textit{D. Performance metrics}. We advance the notation of our final metric in Eq.~\eqref{eq:joint_metric}. We use this notation to condense the deployment problem in the form of Problem~\ref{problem:joint_optimization}. In subsection \textit{C. \name{} in Action} we explain how we intend to use our metric to guide a straightforward solution search method to obtain good quality solutions, and in subsection \textit{D. Performance metrics} we explain how our metric conforms a direct representation of the communications and sensing capabilities of an infrastructure. With this objective, and to make Problem~\ref{problem:joint_optimization} tractable, we leverage on multiple {Q-learning} \gls{rl} agents to explore the feasible solutions, as shown in Fig.~\ref{fig:blocks} and Alg.~\ref{alg:blocks_alg}. The foundations of Q-learning we employ for this are explained in this subsection, \change{ in the \textbf{Reinforcement learning} part}. To employ this technique, we translate the set of possible solutions to Problem~\ref{problem:joint_optimization} into a space of states. As explained \change{in the \textbf{Reinforcement learning} part}, the Q-learning agent will then navigate this space of states by taking actions. Additionally, to guide the agent training we devise our localization and communication metrics, and a way to combine them in different degrees to produce a single, final metric that serves as feedback, evaluating the quality of each solution found, and by extension the quality of each action taken by the agent. These metrics are built by leveraging two key concepts: $i$) the Fisher information and $ii$) the Logistic function.

We would like to emphasize that Q-learning is chosen for our architecture because the discrete nature of the problem formulation makes it adequate for our situation, for its adaptability and guaranteed convergence, but also because its simplicity. More complex solutions employing deep reinforcement learning could be employed, having shown in previous works as~\cite{encinas2023unlocking} ample capacity to solve the deployment problem and interesting capabilities (as better scalability). However they come at a cost that we expressly avoid: a more convoluted architecture.

{\bf Reinforcement learning.} \gls{ris} deployment problems often are, when no simplifications or approximations are employed, NP-Hard~\cite{encinas2024cost}. Traditional tools to create deployment methods rely on relaxations and adaptations of the problem to make it tractable, or solutions are found using approximate methods. While these approaches work, the alternative option of using \gls{ml} methods, and in particular \gls{rl}, allows us to tackle the problem with as much realism as the simulator employed allows for~\cite{encinas2024cost, encinas2023unlocking}. To showcase how our metric can lead a simple \gls{rl} agent to solve the problem at hand, we opt for Q-learning. While there is no dynamic situation for the agent to handle, we create an environment where the agent continuously alters the deployment solution to maximize its performance.

Q-learning is a model-free \gls{rl} algorithm. It learns to optimize the reward obtained by actions in an environment by elaborating a table (called the Q-table) with one entry for every possible state and action, based on the past rewards received. For \name{} we build on the traditional Q-learning architecture, as introduced originally in~\cite{watkins1989learning, watkins1992q}, and schematically presented in Algoritm~\ref{alg:vanilla-q}. On it, we denote the discount factor by $\gamma$, which governs the ability of the agent to take future expected rewards into consideration, instead of following a greedy approach. 

\begin{algorithm}
\caption{Q-Learning Algorithm}
\label{alg:vanilla-q}
\begin{algorithmic}[1]
\State \textbf{Initialize} Q-table arbitrarily (e.g., with zeros) for all state-action pairs
\State \textbf{Set} parameters: learning rate \( \alpha \), discount factor \( \gamma \), exploration rate \( \epsilon \)
\For{each episode}
    \State \textbf{Initialize} state \( s_0 \)
    \For{each step in the episode}
        \State Choose action \( a_t \) based on the exploration-exploitation strategy
        \State Take action \( a_t \), observe reward \( r_{t+1} \), and transition to next state \( s_{t+1} \)
        \State \textbf{Update} Q-table:
        \begin{align*}
        Q(s_t, a_t) \gets &(1-\alpha) \cdot Q(s_t, a_t)  \\
        \quad &+ \alpha \left( r_{t+1} + \gamma \max_{a} Q(s_{t+1}, a) \right)
        \end{align*}
        \State \textbf{Set} \( s_t \gets s_{t+1} \)
    \EndFor
\EndFor
\end{algorithmic}
\end{algorithm}

\begin{algorithm}
\caption{\change{Overview of \name{} in pseudocode form.}}
\label{alg:blocks_alg}
\begin{algorithmic}[1]
\State \change{\textbf{Initialize} solution evaluation memory}
\While{\change{Total training time not exhausted}}
    \State \change{\textbf{Instantiate} Q-learning agent}
    \While{\change{Individual training time not exhausted}}
    \State \change{Train Q-learning agent (Alg.\ref{alg:vanilla-q})}
    \If{\change{Solution state $s_t$ or $s_{t+1}$ exist in memory}}
        \State \change{Fetch solution evaluation from memory}
    \Else
        \State \change{Evaluate $ \frac{1}{A_{u}} \ \underset{\mathcal{A}_{u}}{{\iint}} M(\bold{d},x_{u},y_{u}) \, dx \ dy$}
        \State \change{Store solution evaluation in memory}
    \EndIf
    \If{\change{Budget $\kappa(\bold{d}) > \beta$}}
        \State $s_{t+1} \gets s_t$
        \State $r_{t+1} \gets 0$
    \EndIf
    \State \change{\textbf{Update} Q-table (Alg.\ref{alg:vanilla-q})}
\EndWhile
\EndWhile
\State \change{Select solution with best performance from solution evaluation memory within budget $\kappa(\bold{d}) \leq \beta$}
\end{algorithmic}
\end{algorithm}

\section{\name{} framework}
\label{sec:solution}

The rationale behind our choice relies on the fact that Q-learning is proven to converge to the optimal solution~\cite{watkins1992q}. In our specific problem we encode each possible deployment solution as an state $s$. Any deployment solution can be encoded in a vector with $n$ positions, one for each \gls{cs} in our scenario. For each position of the vector (each \gls{cs}) we have $m$ mutually exclusive deployment choices: leaving the site empty and $m-1$ different options of devices to install. That way, the total number of possible solutions is $m^n$. Since the set of possible solutions is finite, we establish a bijective relation between each solution vector and a state number. The set of actions we define for the agent to navigate in this space is composed of $2\cdot n$ options: for each \gls{cs} in the vector the agent can choose to increase or decrease the index of the device installed on that site, being able to empty it. This leaves us with a Q-table with $m^n$ rows and $2\cdot n$ columns.

For the training, we reward each action of the agent with the difference in the metric valuation between the resulting state and the original state of each action. That way, the total-reward convergence of the Q-learning agent~\cite{watkins1992q} warrants the convergence to the optimal solution. We discard the actions that would produce a state which would exceed the given budget. In those cases, we do not change the current state and we feed the agent a zero reward.

The balance between exploitation and exploration in the training action decisions follows a linear evolution, from all-exploration at the beginning to all-exploitation at the end.

{\bf Logistic function.}  We face several challenges while building both our proposed deployment method \name{}, its system model and formal framework. One particularly recurring and central obstacle is the adequate handling of extreme cases in both the localization and communication metrics. Such extreme cases may range between areas (e.g., very close to a \glspl{bs} and \glspl{ris}) that attain location and communication performances several orders of magnitude above others, and far-away areas where services simply do not reach a usable threshold. To encode increases and decreases in already outlying values while ensuring tractability of the resulting performance indicators using data driven and \gls{ml} tools, we use sigmoid functions as a way to preprocess raw values. Among sigmoid functions, we opt for the logistic function to exploit the convenient properties it shows (e.g., symmetry) to make mathematical tractability easier. 
 We then build the metrics here presented using said logistic function, advanced in Section~\ref{sec:system_model} as part of Eq.~\eqref{eq:sensor_var}, with the following form
\begin{equation} \label{eq:logistic}
g(x) = \frac{e^{k(x-x_0)}}{1+e^{k(x-x_0)}}.
\end{equation}
By carefully selecting the parameters $k,x_0$ in the function, we can choose the thresholds between negligible function values and saturated output. This behaviour leads to an optimal use of both communication and localization services as described hereafter. Other functions could serve in these objectives, e.g., saturating the values with maximum and minimum thresholds $\tau_{max}$ and $\tau_{min}$, as $g_s(x) = \max(\tau_{min}, \min(\tau_{max}, x))$, but the logistic function favours its usage for \gls{ml} training efforts being smooth for all $ x \in \mathbb{R}$.

{\bf Fisher information.} The Fisher information is a metric to evaluate the amount of information we can obtain about an unknown parameter $q$ from an observable random variable $Z$. In our case, we want to know how much information we can obtain about the $\mathbb{R}^4$ parameter comprising the original \gls{ue} position and internal clock deviation 
$\textbf{q} = (\textbf{p}_{u}, \Delta_{t_{u}}) = (x_{u},y_{u},z_{u}, \Delta_{t_{u}})$ 
from all our sensor measurements combined, translated to a single \gls{pdf}. 

When considering a parameter $\textbf{q}$, estimated based on an observation vector $Z$, which has a \gls{pdf} $P(Z)$, the Fisher information takes the form of a symmetric square matrix with the same dimension as $\textbf{q}$, with the element in the $i$-th row and $j$-th column defined \cite{Zhao_Guibas_2003} as

\begin{equation} \label{eq:fisher_info}
\mathcal{I}(\textbf{q})_{i, j} =
\operatorname{\mathbb{E}}\left[
\left(\frac{\partial \ln P(Z)}{\partial {q}_i} \right)
\left(\frac{\partial \ln P(Z)}{\partial {q}_j} \right)
\right].
\end{equation}

Our specific application of the Fisher information is detailed later in this section, subsection \textit{D. Performance metrics}, \change{in the \textbf{Localization} part}. Essentially, we employ the statistical distribution of measurements we derive from the sensor models we propose to compute the Fisher information. Numerically, this direct application implies the computation of the \gls{pdf} in a fine grid, obtaining the logarithm in each point of the grid, and the exhaustive computation of Eq.~\eqref{eq:fisher_info}.

The \gls{crb} states that the Fisher information is the upper bound of the precision of any unbiased estimator. Therefore, we use the Fisher information to evaluate each sensor network. Founding \name{} in the Fisher information we can avoid the assumption of a particular localization framework, and hence forgo the evaluation of the localization performance using that method. Instead, we evaluate the amount of information available from our measurement set about the \gls{ue} coordinates.

\subsection{\name{} in Action}
We propose an iterative method for installing and configuring \glspl{ris} to jointly improve localization and communication performances, namely \name{}. It works in cycles, where we take random deployments and alter them step by step using a \gls{rl} agent. The iterative deployment process using the agent does not follow a greedy policy: by rewarding the agent with the difference between the performance metrics of the current and the previous scenario, we train the agent to maximize the performance of the final solution, not the intermediate steps. While training the agents is a fast process, the different deployment evaluations take a lot of computational effort. We obtain the best results by storing these results in a solution evaluation memory (see Fig.~\ref{fig:blocks} and Alg.~\ref{alg:blocks_alg}), and run over it a series of Q-learning agents, which can request any solution evaluation, buffered or not. By running agents that focus their starting random deployments on the best scenarios previously found, we are able to introduce a refining phase for the solutions, which in practice greatly reduces the time needed to reach a certain level of solution performance.

{\bf Solution steps.} The high-level descriptions of the solution stages are $i$) considering a random scenario, we evaluate it with the localization and communication metric described in Section~\ref{sec:metrics}, Eq.~\eqref{eq:M_d}, 
$ii$) altering the scenario using a \gls{rl} agent which attempts to maximize said metric, and $iii$) deciding if the process continues (going back to the step $i$) based on the quality of the current solution and the improvement achieved in the last iterations\,\footnote{Note that the solving algorithm is built as a \gls{ml} training process for a \gls{rl} agent, but the successful production of a deploying-capable \gls{rl} agent is neither necessary nor the objective of the solution process.}. The training is simply exploited as a searching method, refining the solution by attempting a balanced mix of informed changes based on the knowledge acquired from the quality of past probed solutions and trying unexplored options. Hence, the success of this algorithm at finding a good deployment is independent of the success of the training phase of the agent. This gives us the capability of tackling the problem in situations where a traditional approach would fail, either for the requirement of much larger computational resources or because of the simple unfeasibility of it.

{\bf Proposed localization framework.} We make heavy use of the Fisher information to make our results independent from the particular localization method used to infer the \gls{ue} location from the sensor measurements. Nevertheless, the same ideas of representing one-dimensional measurements in the full-dimensional state space where we want to locate the user can be exploited for a localization that takes into account all information at hand. In equations Eq.~\eqref{eq:AoA_PDF},~\eqref{eq:ToA_PDF} and related ones the \gls{pdf} represents the probability of \textit{obtaining} a measurement \textit{given} a true state of the user and the sensor accuracy, but we can use it to create a localization method: given a measurement and the sensor accuracy, we can reconstruct the \gls{pdf} of the true \gls{ue} locations, and combine all sensors information in the same manner described in Eq.~\eqref{eq:prod_PDF}.

With this method we are able to integrate \gls{nlos} readings if the \glspl{pdf} are built taking it into account, for example by changing the simple Gaussian distribution of the measurements to the convolution of that Gaussian and a flat probability of producing a \gls{nlos} reading (a \textit{bias}) in a range, as proposed in Eq.~(9)/Fig.~2 of \cite{jourdan2008position}. This method, while being computationally heavy, can integrate measurements from wildly heterogeneous sources of information, and seamlessly combines all the measurements from the sensors without loss of information neither on the measured value nor in the variances and covariances of the final distribution. While we do not showcase the results of this method in the present paper, we have used it to introduce an additional measurement improvement, explained in the following. 

{\bf Improved location probability distributions.} Building on the former localization framework, we propose a method to discard \gls{nlos} readings for localization. By including an additional step in the localization accuracy measurement we can filter the \gls{nlos} measurements. When we have obtained the combined \gls{pdf} of all the available measurements, the overlap of that total and the individual measurements, defined as the scalar product between the two functions $\left\langle u , v \right\rangle = \underset{\mathbb{R}^2}{\iint} u(x,y) v(x,y) \, dx \ dy$ can serve as a guide to identify the (probable) \gls{nlos} measurements. By discarding those measurements, either \gls{aoa} or \gls{toa} or others that do not overlap with the estimation, we can refine the initial measurement. The convolution of the original measurements with the random \gls{nlos} bias can be discarded, and the original process can be reproduced with the \gls{los} measurements, to produce a more precise result.

\subsection{Performance metrics}
\label{sec:metrics}

The core process within \name{} is to explore the solution space while attempting to maximize our metric (Eq.~\eqref{eq:joint_metric}).
To guide the exploration we reward each action of the agent with the difference between the valuation of the initial scenario (before the action was taken) and the resulting scenario (after the action is taken).
The quality of the solutions found will always have the quality of the metric they rely on as a limiting factor. This makes the performance metrics a point of major importance in our proposal.

To jointly optimize localization and communication performances, we build a joint metric by combining two individual metrics, as shown in Eq.~\eqref{eq:joint_metric}, both adequately evaluating the solutions. 

{\bf Communication.} Our metric $C(\bold{d},x_{u},y_{u})$ for the communication performance can be considered the simplest of the two and it is based on the \gls{snr} that each deployment $\bold{d}$ can provide for a fictitious \gls{ue} placed on coordinates $x_{u},y_{u}$ on the considered surface.
To obtain the \gls{snr} in that point we first model the present \glspl{ris} as signal sources. We compute the different amounts of power received by one antenna element of the $i$-th \gls{ris} (which contains $ne_{i}$ elements) and choose the \gls{bs} that shows the highest power transmission to the \gls{ris}: ${P_{ei} = \underset{j}{max}(P_{j-i})}$, where $P_{ei}$ is the maximum power received by an antenna element of the $i$-th \gls{ris} and $P_{j-i}$ is the power received by one antenna element of the $i$-th \gls{ris} from the $j$-th \gls{bs}.
We multiply that power by the number of elements in the \gls{ris} to obtain the maximum signal power the \gls{ris} is able to reflect, $P_{i} = P_{ei} \cdot ne_{i}$. Knowing the distribution of elements of the \gls{ris} and treating it as a \gls{ura}~\cite{shao2022reconfigurable} able to focus on each direction, we compute the maximum attainable gain by the \gls{ris} for every \gls{aoa} of the signals, $G_i(\theta)$. For the \gls{ue} in location $(x_{u},y_{u})$ the angle between the \gls{ris} and the user is $\theta_{i-u} = \text{atan}2(y_{u}-y_{si}, x_{u}-x_{si}) - \theta_{i}$, where $(x_{si}, y_{si})$ is the location of the $i$-th \gls{ris} and $\theta_{i}$ is its orientation with respect to the reference direction.
Having the maximum power that the \gls{ris} can reflect $P_{i}$ and the gain it shows in the direction of the \gls{ue}, and assuming an isotropic antenna in the \gls{ue}, we simply apply Eq.~\eqref{eq:friis} to obtain the received power by the \gls{ue}~\cite{shao2022reconfigurable}. We obtain the \gls{snr} by dividing this amount by a background noise level. Considering the values of the \gls{snr} received from all the present signal emitters (both \glspl{ris} and \glspl{bs}), we use the highest one as the \gls{snr} available for the \gls{ue} $\text{SNR}_u = \underset{k}{max}(\text{SNR}_{u,k})$, where $\text{SNR}_{u,k}$ is the \gls{snr} received by a \gls{ue} at the position $x_{u},y_{u}$ from the $k$-th signal emitter. To turn this value into a score between $0$ and $1$, we apply the logistic function Eq.~\eqref{eq:logistic}. To choose the parameters $k,x_0$ we take as milestone values the expected minimum threshold at which mobile radio connectivity can be provided in \gls{b5g}, and the expected maximum threshold beyond which no noticeable further improvement is obtained, i.e., $\text{SNR}_{min}$ and $\text{SNR}_{max}$, accordingly. We finally have the following

\begin{equation}
C(\bold{d},x_{u},y_{u}) = \frac{e^{k_{\text{SNR}}(\text{SNR}_u-\text{SNR}_c)}}{1+e^{k_{\text{SNR}}(\text{SNR}_u-\text{SNR}_c)}}.\label{eq:C_dxy}
\end{equation}
Since we want this function to return a value of $0.9$ when $\text{SNR}_u = \text{SNR}_{max}$, and by $0.1$ when $\text{SNR}_u = \text{SNR}_{min}$, we pick $k_{\text{SNR}}$ and $\text{SNR}_c$ as per Eq.~\eqref{eq:k_SNR} and Eq.~\eqref{eq:SNR_c}.

To turn this metric $C(\bold{d},x_{u},y_{u})$, which evaluates a single point, into the scenario-wide metric $C_S(\bold{d})$ that assesses the deployment $\bold{d}$ over the considered area $\mathcal{A}_{u}$, we simply take and normalize the integral over $\mathcal{A}_u$ by averaging the value of the metric as follows

\begin{equation} \label{eq:C_d}
C_S(\bold{d}) = \frac{1}{A_{u}} \ \underset{\mathcal{A}_{u}}{{\iint}} C(\bold{d},x_{u},y_{u}) \, dx \ dy.
\end{equation}

{\bf Localization.} Our metric $L(\bold{d},x_{u},y_{u})$ for the localization performance is based on the Fisher information of the \gls{pdf} of the \gls{ue} location measurements. We first consider all the possible measurement distributions for a fictitious \gls{ue} placed on coordinates $(x_{u},y_{u})$ on the considered surface, from all the available sensors in the scenario. We then obtain the measured \gls{ue} position \glspl{pdf}. These take the form of Eq.~\eqref{eq:AoA_PDF} for \gls{aoa} measurements and Eq.~\eqref{eq:ToA_PDF} for \gls{toa} measurements. We condense them in a single \gls{pdf} $P_R(x,y)$ following Eq.~\eqref{eq:prod_PDF}:
this digests all the positional available information from the sensors in the scenario. Hence, we compute its Fisher information matrix, with the parameter $\bold{q} = (x,y)$ in Eq.~\eqref{eq:fisher_info} where $q_1 = x$ and $q_2 = y$, it yields 
\begin{equation}
\mathcal{I}(\textbf{q}|x_{u},y_{u})_{i, j}\!=\!
\operatorname{\mathbb{E}}\!\left[\!
\left(\frac{\partial \ln P_R(x,y)}{\partial {q}_i} \right)\!\!
\left(\frac{\partial \ln P_R(x,y)}{\partial {q}_j} \right)\!
\right]\!.
\end{equation}

In practice, as introduced \change{in the \textbf{Fisher information} part} of subsection \textit{B. \name{} Architecture}, we solve this equation numerically. To do this, we compute $P_R(x,y)$ in a fine grid, obtain the logarithm of each value of the grid, make the numerical derivatives in ${q}_1 = x$ and ${q}_2 = y$, and obtain the expected value in the formula through the product with the original distribution $P_R(x,y)$ and numerical integration.

The Fisher information matrix of $P_R(x,y)$ provides $3$ distinct values, $\mathcal{I}(\bold{q})_{1,1}$, $\mathcal{I}(\bold{q})_{2,2}$ and $\mathcal{I}(\bold{q})_{1,2} = \mathcal{I}(\bold{q})_{2,1}$. Since the localization performance is ultimately limited by the minimum localization accuracy in any direction, and we want to make our metric independent under rotations, we summarize the information matrix into a single value, taking the smallest of the eigenvalues of the matrix. We diagonalise the Fisher information matrix as $\mathcal{I}(x, y) = {P}\mathcal{I}_D{P}^{-1}$. We take the smallest value of the diagonal in the new base as the minimum spatial information in any direction, $\mathcal{I}_{min} = min\{ev_1(\mathcal{I}), ev_2(\mathcal{I})\}$.

Leveraging on the \gls{crb}, we compute the inverse of the square root to obtain the minimum bound of the standard deviation any unbiased position estimator would be able to achieve when locating the \glspl{ue} in the direction of greatest uncertainty, $\sigma_{\text{CRB},u} = 1/\sqrt{\mathcal{I}_{min}}$, using the deployment of sensing infrastructure $\bold{d}$ initially given.
Localization services are usually required to meet a sharp threshold. Increased accuracy above this threshold brings no additional benefit, and performance falling short of it renders the service valueless. To model this, given a localization accuracy threshold $\sigma_{\text{TH}}$, we resort again to the logistic function Eq.~\eqref{eq:logistic} (but inverted, since we want to reward low $\sigma_{\text{CRB},u}$ values) to obtain our final metric as the following:

\begin{equation}
L(\bold{d},x_{u},y_{u}) = \frac{1+e^{k_{\text{CRB}}(\sigma_{\text{CRB},u}-\sigma_{\text{TH}})}}{e^{k_{\text{CRB}}(\sigma_{\text{CRB},u}-\sigma_{\text{TH}})}}.\label{eq:L_dxy}
\end{equation}

In this case, the threshold value $\sigma_{\text{TH}}$ has to be chosen as the localization accuracy that our service aims to provide, and $k_{\text{CRB}}$ needs to be high enough to render the logistic function a soft step, but keeping a slope that enables \gls{ml} methods. \change{When using meters for the $\sigma_{\text{CRB},u}$ values, a value of $k_{\text{CRB}} = 10 \cdot ln 9 \approx 21.97$ provides a change from $0.1$ to $0.9$ when going from $\sigma_{\text{TH}}-10cm$ to $\sigma_{\text{TH}}+10cm$ and a central slope of $5m^{-1}$. In our numerical evaluation we use a $\sigma_{\text{TH}} = 1$m.}

As with the previous metric, to turn this metric $L(\bold{d},x_{u},y_{u})$, which evaluates a single point, into the scenario-wide metric $L_S(\bold{d})$, which evaluates the deployment $\bold{d}$ over the whole area $\mathcal{A}_{u}$, we simply take and normalize the integral over $\mathcal{A}_u$, averaging the value of the metric:

\begin{equation} \label{eq:L_d}
L_S(\bold{d}) = \frac{1}{A_{u}} \ \underset{\mathcal{A}_{u}}{{\iint}} L(\bold{d},x_{u},y_{u}) \, dx \ dy.
\end{equation}

Naturally, we extend Eq.~\eqref{eq:joint_metric} to the scenario-wide metrics Eq.~\eqref{eq:C_d} and  Eq.~\eqref{eq:L_d} as follows:

\begin{equation} \label{eq:M_d}
M_S(\bold{d}) = \alpha \ L_S(\bold{d}) + (1-\alpha) \ C_S(\bold{d}).
\end{equation}

{\bf Budget.} An additional external factor we take into account for deployments is the budget constraint. This energy or cost limitation takes the form of a set of rules that can evaluate the resources required by any proposed solution and rule it viable or not viable. The variables included in this budget evaluation are the total available budget and any relevant \gls{ris} or \gls{bs} characteristics, such as size, number of elements, sensing capabilities, or location. The budget rule acts simply by discarding unfeasible solutions, so it does not need to follow any linearity or derivability conditions, simply to deem a deployment doable or not.

\section{Numerical Evaluation}
\label{sec:results}
We evaluate \name{} by means of exhaustive numerical simulations using a commercial tool, namely MATLAB.

\subsection{Methodology and simulation specifics}
We use discrete probability distributions as homogeneous samplings of the corresponding continuous form: the value of the individual samples does not change with the spatial frequency of the sampling. If a two-dimensional \gls{pdf} $P$ is sampled within an area in the points $\textbf{p}_{i,j} = (x_i, y_j)$ where $\Delta_x = x_{i} - x_{i-1} \ \forall{i>1}$ and $\Delta_y = y_{j} - y_{j-1} \ \forall{j>1}$, it is normalised as ${\sum_{i,j}^{} P_{i,j} \cdot \Delta_x \cdot \Delta_y = 1}$, where $P_{i,j}$ is the $i,j$-th sample of the \gls{pdf} $P$ and $\Delta_x, \Delta_y$ are the spatial granularity of the sampling. We use a spatial granularity of $1$m.

In addition, we assume the values for $\text{SNR}_{min}$ and $\text{SNR}_{max}$ according to current technological typical values at $0$dB and $20$dB respectively, based on~\cite{ETSI38_104}. \change{As per localization, we take an accuracy threshold of $\sigma_{\text{TH}} = 1$m as a conservative reference for near future networks~\cite{liu2017prospective}, and $\gamma = 2$.}

\begin{figure}[t]
        \center
      \includegraphics[width=0.7\linewidth, trim = {0cm 0cm 0cm 0cm}]{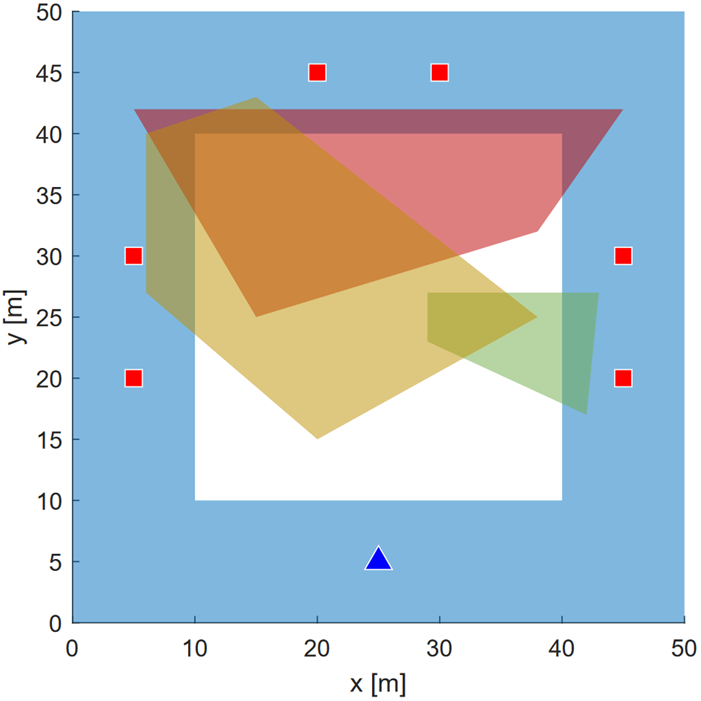} 
        \caption{Example of a base scenario, with CSs, one BS and shadowing areas.}
        \label{fig:scenario_areas}
\end{figure}
 
The scenario for the numerical evaluation is depicted in Fig.~\ref{fig:scenario_areas}. On it, we can see, from a zenithal view, the complete playground forming a $50m \times 50m$ square in the $z=0m$ plane, from the origin to the $x = 50m , y = 50m$ point. In the central area, centered in the scenario, measuring $30m \times 30m$, and painted white in Fig.~\ref{fig:scenario_areas}, we distribute homogeneously the \glspl{ue}. We use these \glspl{ue} to probe the capabilities of the possible deployments of devices. This \gls{ue} area goes from $x = 10m , y = 10m$ to $x = 40m , y = 40m$. In the bottom of Fig.~\ref{fig:scenario_areas} we can see where we place the \gls{bs} we consider in the simulation, at $x = 25m, y = 5m$. In this example we consider $6$ possible places where a \gls{ris} can be placed, i.e., the \glspl{cs}, and we represent them as red squares. The shaded regions in red, mustard, and green represent the shadow areas that affect the \gls{bs} and \glspl{cs}.
The moderate number of \glspl{cs} allows us to realistically simulate a medium-sized scenario while avoiding the steep growth of computational costs associated with the NP-Hardness of the problem~\cite{nguyen2018mobile}. We operate at a frequency of $f = 24$GHz with a bandwidth of $B = 100$MHz for the \gls{toa} measurements. We assume a background noise of $-80$dBm whereas we have two available deployable \gls{ris} models, one with $40\times10$ elements and one with $80\times20$. The budget rule values these at $
\beta_{1} = 100\$$ and $\beta_{2} = 400\$$ respectively with an install cost of $\beta_{ins} = 200\$$. We consider a total budget $\beta_{tot} = 1200\$$.\footnote{For the \gls{ris} cost estimation we take into account the number of elements of the considered devices, the frequency of $f = 24$GHz, an inter-element distance of $\lambda/2$, current costs for large batches of large-scale printed circuit boards, and an estimation of the costs per element in existing prototypes. Nevertheless, the exact values are not central to our results, as the relevant feature we want to showcase is the capability of \name{} to work within a budget and to produce different solutions within that constraint. As specified before, the budget constrain can represent a limit in either energy consumption, economic cost, or both.}

\subsection{Results}

\begin{figure}[t]
        \center
      \includegraphics[width=\linewidth, trim = {4cm 6.3cm 2cm 4cm}]{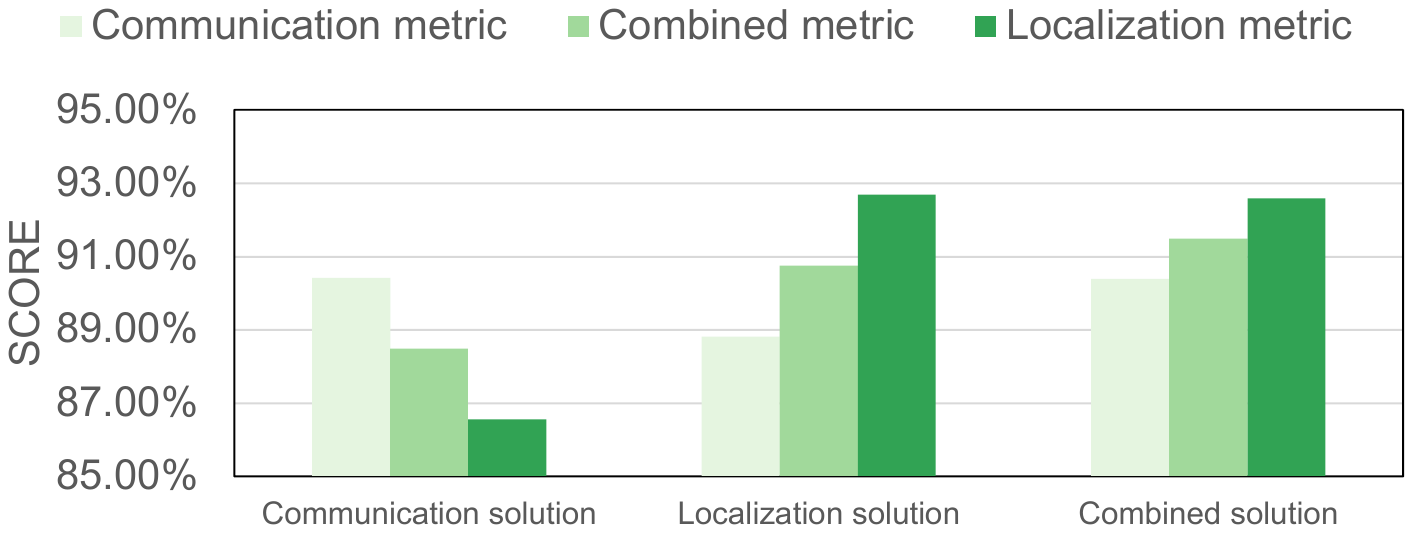} 
        \caption{Performance comparison of different solutions: communications-focused (legacy), localization-focused, and combined \gls{isac} (\name{}) guided.}
        \label{fig:solutions_bar}
\end{figure}

The Q-learning agents, using the buffered solution evaluation and the differential reward, are able to find outstanding results\,\footnote{We have empirically tested that \name{} results are near-optimal by comparing them with the results from exhaustive searches; however, due to the complexity of the problem, this comparison could only be executed for very limited instances of the original problem.}. 
Traditional solutions to the challenge of deploying communication infrastructure typically focus solely on communication metrics. In our study, we obtain similar solutions to serve as benchmarks by running our solver with communication metrics as the only form of feedback, i.e. Eq.~\eqref{eq:M_d} with $\alpha = 0$.
On the other hand, the---more scarce---sensor deployments only take into consideration the accuracy of their combined information. We can observe how the best results obtained by looking only at one specific metric according to Eq.~\eqref{eq:M_d} (with $\alpha=0$ in case of communication-oriented metric and $\alpha=1$ for localization-oriented metric), often show very poor performance towards the neglected metric, as suggested by Fig.~\ref{fig:solutions_bar}. In other words, the trade-off of these two metrics is made apparent in said Fig.~\ref{fig:solutions_bar}. 
It is clear that a single-objective solution search can result in a slight improvement in their metric of focus while significantly degrading the quality of the deployment in the other.

\begin{figure}[t]
        \center
      \includegraphics[width=\linewidth, trim = {4cm 6cm 2cm 5cm}]{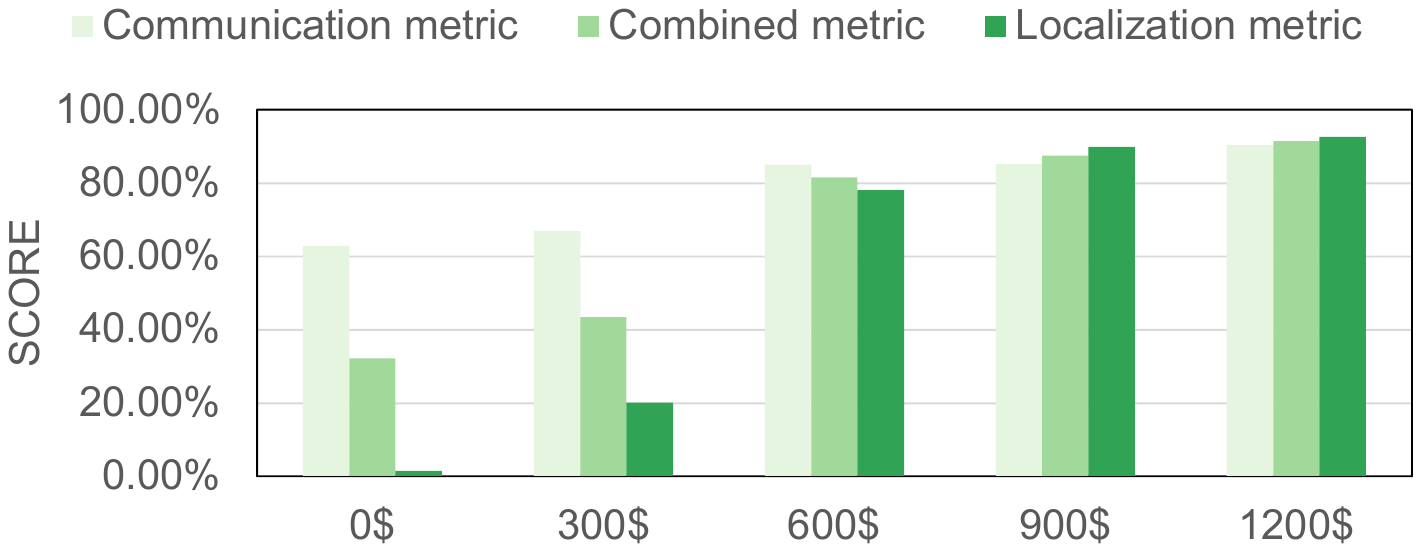} 
        \caption{Performance comparison of solutions for different budgets.}
        \label{fig:budgets_bar}
\end{figure}

Indeed, the necessity of a proper localization metric can be further noticed in Fig.~\ref{fig:budgets_bar}: Solutions with reasonable communication performance often lead to unacceptable localization accuracy. Furthermore, the usage of the combined metric has proven highly successful: the corresponding solutions usually exhibit at the same time almost-optimal performance in both communication and localization (Fig.~\ref{fig:solutions_bar}). Such solutions often remain hidden when we consider communication or localization-oriented performance alone.

In Fig.~\ref{fig:budgets_bar} we can observe one of the side results of the solution buffer: we can analyze the best solutions when different budget rules are applied. This breakdown of the results can greatly help the determination of optimized deployments in infrastructure within economic or energy efficiency considerations. In the example, we can see how a budget limit $\beta$ of half of the initially allocated amount can readily achieve almost the same performance as the best solution with the full resource expenditure. We can also compare deployments against the current infrastructure performance in both metrics used as baselines. 
Our presented solution can be easily expanded if provided with a realistic translation able to map performance gains to economic gains, taking into account projected network usage given additional coverage or throughput, projected market share under the improved performance, etc. With such information, the metrics and the solution that we propose in this paper can easily guide future deployments economic and energy efficiency terms, focused either on the optimization of the benefits obtained per consumed watt or the maximum \gls{roi}.

\begin{figure}[t]
        \center 
      \includegraphics[width=\linewidth, trim = {0cm .0cm 0cm .0cm}]{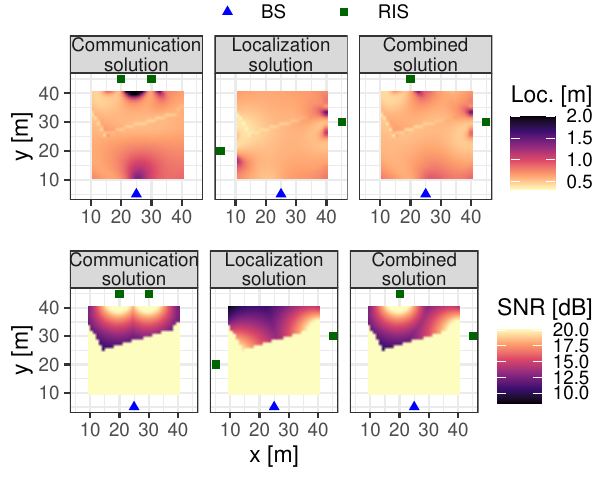} 
        \caption{Localization and communication metric maps of the communications-focused, localization-focused, and combined metric-guided solutions.}
        \label{fig:maps}
\end{figure}

As explained in Sec.~\ref{sec:solution}, the core of our proposal, \name{}, is on the metrics used to guide the solution-finding process. We use Fig.~\ref{fig:maps} to show at the same time the aspect of these metrics, and the differences between the different deployments we find using \name{}. In Fig.~\ref{fig:maps} we have three columns (one for each solution found for the scenario) and two rows (one for each metric). The three solutions are those already represented in Fig.~\ref{fig:solutions_bar} and Fig.~\ref{fig:budgets_bar}: one that maximizes the communication metric, one that maximizes the localization metric, and the last one, maximizing the combined metric. 
Regarding the rows, we devote the upper one to the localization metric: in particular, we plot $\sigma_{\text{CRB},u} = 1/\sqrt{\mathcal{I}_{min}}$, the core of the localization metric. This represents the maximum achievable accuracy that each deployment solution is able to reach across the simulated scenario $\mathcal{A}_u$.
This value is chosen for its intuitive meaning: To use it, as detailed in Sec.~\ref{sec:solution}, we map it to the $(0,1)$ interval with the logistic function to form Eq.~\eqref{eq:L_dxy}, and then integrate it across the area $\mathcal{A}_u$ to obtain Eq.~\eqref{eq:L_d}.
In the lower row we represent the coverage metric. Similarly, we represent \gls{snr}, which process using the logistic function to obtain Eq.~\eqref{eq:C_dxy} and integrate in Eq.~\eqref{eq:C_d}.
With this we illustrate at the same time two important parts of this work: on one hand, how the localization metric measures the accuracy achievable by combining the information available from present sensors, and on the other hand, how we combine that with the communications metric. Fig.~\ref{fig:maps} serves to visualize how the best solutions found to satisfy each metric can diverge in shape.

In Fig.~\ref{fig:maps} we can see too where within $\mathcal{A}_u$ we meet or fall short of our communication and localization targets. We can see how the deployment greatly enhances communication in all cases, almost eliminating dead zones, exhibiting a minimum \gls{snr} of $\approx7.5$dB, way above $\text{SNR}_{min} = 0$dB. Notably, the experienced localization accuracy, not available at a single-BS infrastructure, hits the demanding performance of $1$m in large spatial extensions in the localization and combined solutions.

While a formal, thorough complexity analysis would be impractical due to the random nature of Q-learning, we have observed that the time $t_c$ needed for the convergence of the solutions shows an approximate linear dependence with the maximum number of devices that can be installed for the allocated budget $t_{c} \sim 1800s \cdot \beta_{tot} / (\min_i(\beta_i)+\beta_{ins})$, for an approximate time of two hours in the runs detailed here, running on a personal computer equipped with a 12th Gen Intel\textregistered\space Core\texttrademark\space i7 and $32$ GBytes of random access memory.

\section{Conclusions}
This paper highlights the transformative potential of \gls{ris} in powering the sensing and communications revolution within the context of 6G networks. \glspl{ris} offer unprecedented opportunities to enhance the performance and efficiency of future wireless networks, enabling a new use of connectivity and data exchange. In particular, we have studied how both localization and communication-oriented design can coexist in wireless networks by proposing an evaluation metric. Based on it, we pioneer \name{}, a novel framework that optimally designs \gls{ris} placement, achieving with them an outstanding, next-gen localization performance attaining sub-meter accuracy, improving over traditional approaches focused either on communications-only or accuracy-only metrics. 

Obtained results and the demonstrated flexibility and adaptability of our novel architecture will open the door for future work, for example, revolving around the usage of ray tracing simulation tools or more in-depth economic conditions for real-life commercial applications.

\section{Acknowledgement}

This work was partially supported by the SNS JU Horizon Europe project under Grant Agreement 101139130 (6G-DISAC), and by the SNS JU Horizon Europe Project under Grant Agreement No. 101192521 (MultiX), and by the European Union under the Italian National Recovery and Resilience Plan (NRRP) of NextGenerationEU, partnership on “Telecommunications of the Future” (PE00000001 - program “RESTART”) – CUP E63C22002040007, and is part of the Grant 6G-DIFERENTE (CER-20231018) funded by CDTI and by the European Union NextGenerationEU/PRTR within the call "Ayudas Cervera para Centros Tecnológicos 2023"

\bibliographystyle{IEEEtran}
\bibliography{references}

\begin{thebibliography}{10}
\providecommand{\url}[1]{#1}
\csname url@samestyle\endcsname
\providecommand{\newblock}{\relax}
\providecommand{\bibinfo}[2]{#2}
\providecommand{\BIBentrySTDinterwordspacing}{\spaceskip=0pt\relax}
\providecommand{\BIBentryALTinterwordstretchfactor}{4}
\providecommand{\BIBentryALTinterwordspacing}{\spaceskip=\fontdimen2\font plus
\BIBentryALTinterwordstretchfactor\fontdimen3\font minus
  \fontdimen4\font\relax}
\providecommand{\BIBforeignlanguage}[2]{{%
\expandafter\ifx\csname l@#1\endcsname\relax
\typeout{** WARNING: IEEEtran.bst: No hyphenation pattern has been}%
\typeout{** loaded for the language `#1'. Using the pattern for}%
\typeout{** the default language instead.}%
\else
\language=\csname l@#1\endcsname
\fi
#2}}
\providecommand{\BIBdecl}{\relax}
\BIBdecl

\bibitem{Liu22_JSAC}
F.~Liu, Y.~Cui, C.~Masouros, J.~Xu, T.~X. Han, Y.~C. Eldar, and S.~Buzzi,
  ``Integrated sensing and communications: Toward dual-functional wireless
  networks for {6G} and beyond,'' \emph{IEEE Journal on Selected Areas in
  Communications}, vol.~40, no.~6, pp. 1728--1767, 2022.

\bibitem{lui20_tcomm}
F.~Liu, C.~Masouros, A.~P. Petropulu, H.~Griffiths, and L.~Hanzo, ``Joint radar
  and communication design: Applications, state-of-the-art, and the road
  ahead,'' \emph{IEEE Transactions on Communications}, vol.~68, no.~6, pp.
  3834--3862, 2020.

\bibitem{zhang22_comst}
J.~A. Zhang, M.~L. Rahman, K.~Wu, X.~Huang, Y.~J. Guo, S.~Chen, and J.~Yuan,
  ``Enabling joint communication and radar sensing in mobile networks—a
  survey,'' \emph{IEEE Communications Surveys \& Tutorials}, vol.~24, no.~1,
  pp. 306--345, 2022.

\bibitem{gruber08_tap}
F.~K. Gruber and E.~A. Marengo, ``New aspects of electromagnetic information
  theory for wireless and antenna systems,'' \emph{IEEE Transactions on
  Antennas and Propagation}, vol.~56, no.~11, pp. 3470--3484, 2008.

\bibitem{Abrardo21}
A.~Abrardo, D.~Dardari, M.~{DI Renzo}, and X.~Qian, ``{MIMO Interference
  Channels Assisted by Reconfigurable Intelligent Surfaces: Mutual Coupling
  Aware Sum-Rate Optimization Based on a Mutual Impedance Channel Model},''
  \emph{IEEE Wireless Communications Letters}, vol.~10, no.~12, pp. 2624--2628,
  2021.

\bibitem{di2020smart}
M.~Di~Renzo, A.~Zappone, M.~Debbah, M.-S. Alouini, C.~Yuen, J.~De~Rosny, and
  S.~Tretyakov, ``Smart radio environments empowered by reconfigurable
  intelligent surfaces: How it works, state of research, and the road ahead,''
  \emph{IEEE journal on selected areas in communications}, vol.~38, no.~11, pp.
  2450--2525, 2020.

\bibitem{albanese2022ris}
A.~Albanese, G.~Encinas-Lago, V.~Sciancalepore, X.~Costa-P{\'e}rez, D.-T.
  Phan-Huy, and S.~Ros, ``{RIS}-aware indoor network planning: The rennes
  railway station case,'' in \emph{ICC 2022-IEEE International Conference on
  Communications}.\hskip 1em plus 0.5em minus 0.4em\relax IEEE, 2022, pp.
  2028--2034.

\bibitem{encinas2023unlocking}
G.~Encinas-Lago, A.~Albanese, V.~Sciancalepore, M.~Di~Renzo, and
  X.~Costa-P{\'e}rez, ``Unlocking metasurface practicality for {B5G} networks:
  {AI}-assisted {RIS} planning,'' in \emph{GLOBECOM 2023-2023 IEEE Global
  Communications Conference}.\hskip 1em plus 0.5em minus 0.4em\relax IEEE,
  2023, pp. 6560--6566.

\bibitem{gan2024coverage}
X.~Gan, C.~Huang, Z.~Yang, X.~Chen, J.~He, Z.~Zhang, C.~Yuen, Y.~L. Guan, and
  M.~Debbah, ``Coverage and rate analysis for integrated sensing and
  communication networks,'' \emph{IEEE Journal on Selected Areas in
  Communications}, 2024.

\bibitem{an2023fundamental}
J.~An, H.~Li, D.~W.~K. Ng, and C.~Yuen, ``Fundamental detection probability vs.
  achievable rate tradeoff in integrated sensing and communication systems,''
  \emph{IEEE Transactions on Wireless Communications}, vol.~22, no.~12, pp.
  9835--9853, 2023.

\bibitem{albanese22_tmc}
A.~Albanese, V.~Sciancalepore, A.~Banchs, and X.~Costa-Perez, ``{LOKO}:
  Localization-aware roll-out planning for future mobile networks,'' \emph{IEEE
  Transactions on Mobile Computing}, pp. 1--1, 2022.

\bibitem{Ma23_ojcoms}
T.~Ma, Y.~Xiao, X.~Lei, L.~Zhang, Y.~Niu, and G.~K. Karagiannidis,
  ``Reconfigurable intelligent surface-assisted localization: Technologies,
  challenges, and the road ahead,'' \emph{IEEE Open Journal of the
  Communications Society}, vol.~4, pp. 1430--1451, 2023.

\bibitem{liu2023integrated}
R.~Liu, M.~Li, H.~Luo, Q.~Liu, and A.~L. Swindlehurst, ``Integrated sensing and
  communication with reconfigurable intelligent surfaces: Opportunities,
  applications, and future directions,'' \emph{IEEE Wireless Communications},
  vol.~30, no.~1, pp. 50--57, 2023.

\bibitem{chen21_twc}
Y.~Chen, M.~Wen, E.~Basar, Y.-C. Wu, L.~Wang, and W.~Liu, ``Exploiting
  reconfigurable intelligent surfaces in edge caching: Joint hybrid beamforming
  and content placement optimization,'' \emph{IEEE Transactions on Wireless
  Communications}, vol.~20, no.~12, pp. 7799--7812, 2021.

\bibitem{zeng21_lcomm}
S.~Zeng, H.~Zhang, B.~Di, Z.~Han, and L.~Song, ``Reconfigurable intelligent
  surface {(RIS)} assisted wireless coverage extension: {RIS} orientation and
  location optimization,'' \emph{IEEE Communications Letters}, vol.~25, no.~1,
  pp. 269--273, 2021.

\bibitem{ntontin21_ojcom}
K.~Ntontin, A.-A.~A. Boulogeorgos, D.~G. Selimis, F.~I. Lazarakis, A.~Alexiou,
  and S.~Chatzinotas, ``Reconfigurable intelligent surface optimal placement in
  millimeter-wave networks,'' \emph{IEEE Open Journal of the Communications
  Society}, vol.~2, pp. 704--718, 2021.

\bibitem{hou22_ictc}
S.~Hou, N.~U. Saqib, S.~H. Chae, Z.~Dou, and S.-W. Jeon, ``Optimal placement of
  reconfigurable intelligent surface for millimeter-wave indoor
  communication,'' in \emph{2022 13th International Conference on Information
  and Communication Technology Convergence (ICTC)}, 2022, pp. 698--700.

\bibitem{ntontin21_eucap}
K.~Ntontin, D.~Selimis, A.-A.~A. Boulogeorgos, A.~Alexandridis, A.~Tsolis,
  V.~Vlachodimitropoulos, and F.~Lazarakis, ``Optimal reconfigurable
  intelligent surface placement in millimeter-wave communications,'' in
  \emph{2021 15th European Conference on Antennas and Propagation (EuCAP)},
  2021, pp. 1--5.

\bibitem{stratidakis22_tap}
G.~Stratidakis, S.~Droulias, and A.~Alexiou, ``Optimal position and orientation
  study of reconfigurable intelligent surfaces in a mobile user environment,''
  \emph{IEEE Transactions on Antennas and Propagation}, vol.~70, no.~10, pp.
  8863--8871, 2022.

\bibitem{lau2017sensor}
B.~P.~L. Lau, N.~Wijerathne, B.~K.~K. Ng, and C.~Yuen, ``Sensor fusion for
  public space utilization monitoring in a smart city,'' \emph{IEEE Internet of
  Things Journal}, vol.~5, no.~2, pp. 473--481, 2017.

\bibitem{wei2022mmwave}
Z.~Wei, F.~Zhang, S.~Chang, Y.~Liu, H.~Wu, and Z.~Feng, ``Mmwave radar and
  vision fusion for object detection in autonomous driving: A review,''
  \emph{Sensors}, vol.~22, no.~7, p. 2542, 2022.

\bibitem{wang2015deepfi}
X.~Wang, L.~Gao, S.~Mao, and S.~Pandey, ``Deepfi: Deep learning for indoor
  fingerprinting using channel state information,'' in \emph{2015 IEEE wireless
  communications and networking conference (WCNC)}.\hskip 1em plus 0.5em minus
  0.4em\relax IEEE, 2015, pp. 1666--1671.

\bibitem{wen2024shaping}
C.-K. Wen, L.-S. Tsai, A.~Shojaeifard, P.-K. Liao, K.-K. Wong, and C.-B. Chae,
  ``Shaping a smarter electromagnetic landscape: {IAB}, {NCR}, and {RIS} in
  {5G} standard and future {6G},'' \emph{IEEE Communications Standards
  Magazine}, vol.~8, no.~1, pp. 72--78, 2024.

\bibitem{zhang2021performance}
Y.~Zhang, J.~Zhang, M.~Di~Renzo, H.~Xiao, and B.~Ai, ``Performance analysis of
  {RIS}-aided systems with practical phase shift and amplitude response,''
  \emph{IEEE Transactions on Vehicular Technology}, vol.~70, no.~5, pp.
  4501--4511, 2021.

\bibitem{Curry_2005}
G.~R. Curry, \emph{Radar System Performance Modeling}.\hskip 1em plus 0.5em
  minus 0.4em\relax Artech House, 2005.

\bibitem{1054120}
M.~Lichtenstein and T.~Young, ``The resolution of closely spaced signals,''
  \emph{IEEE Transactions on Information Theory}, vol.~14, no.~2, pp. 288--293,
  1968.

\bibitem{he2020adaptive}
J.~He, H.~Wymeersch, T.~Sanguanpuak, O.~Silv{\'e}n, and M.~Juntti, ``Adaptive
  beamforming design for mmwave {RIS}-aided joint localization and
  communication,'' in \emph{2020 IEEE Wireless Communications and Networking
  Conference Workshops (WCNCW)}.\hskip 1em plus 0.5em minus 0.4em\relax IEEE,
  2020, pp. 1--6.

\bibitem{keykhosravi2022ris}
K.~Keykhosravi, G.~Seco-Granados, G.~C. Alexandropoulos, and H.~Wymeersch,
  ``{RIS}-enabled self-localization: Leveraging controllable reflections with
  zero access points,'' in \emph{ICC 2022-IEEE International Conference on
  Communications}.\hskip 1em plus 0.5em minus 0.4em\relax IEEE, 2022, pp.
  2852--2857.

\bibitem{xing2021millimeter}
Y.~Xing, T.~S. Rappaport, and A.~Ghosh, ``Millimeter wave and sub-thz indoor
  radio propagation channel measurements, models, and comparisons in an office
  environment,'' \emph{IEEE Communications Letters}, vol.~25, no.~10, pp.
  3151--3155, 2021.

\bibitem{sulyman2014radio}
A.~I. Sulyman, A.~T. Nassar, M.~K. Samimi, G.~R. MacCartney, T.~S. Rappaport,
  and A.~Alsanie, ``Radio propagation path loss models for {5G} cellular
  networks in the 28 {GHz} and 38 {GHz} millimeter-wave bands,'' \emph{IEEE
  communications magazine}, vol.~52, no.~9, pp. 78--86, 2014.

\bibitem{sihlbom2022reconfigurable}
B.~Sihlbom, M.~I. Poulakis, and M.~Di~Renzo, ``Reconfigurable intelligent
  surfaces: Performance assessment through a system-level simulator,''
  \emph{IEEE Wireless Communications}, vol.~30, no.~4, pp. 98--106, 2022.

\bibitem{nguyen2018mobile}
N.-T. Nguyen and B.-H. Liu, ``The mobile sensor deployment problem and the
  target coverage problem in mobile wireless sensor networks are np-hard,''
  \emph{IEEE Systems Journal}, vol.~13, no.~2, pp. 1312--1315, 2018.

\bibitem{encinas2024cost}
G.~Encinas-Lago, A.~Albanese, V.~Sciancalepore, X.~Costa-P{\'e}rez, A.~Banchs,
  and D.-T. Phan-Huy, ``A cost-effective riss deployment to abate the coverage
  problem in b5g networks,'' \emph{IEEE Transactions on Wireless
  Communications}, 2024.

\bibitem{watkins1989learning}
C.~J. C.~H. Watkins, ``Learning from delayed rewards,'' 1989.

\bibitem{watkins1992q}
C.~J. Watkins and P.~Dayan, ``Q-learning,'' \emph{Machine learning}, vol.~8,
  pp. 279--292, 1992.

\bibitem{Zhao_Guibas_2003}
F.~Zhao and L.~J. Guibas, \emph{Information processing in sensor networks:
  Second international workshop, IPSN 2003: Proceedings}.\hskip 1em plus 0.5em
  minus 0.4em\relax Springer, 2003.

\bibitem{jourdan2008position}
D.~B. Jourdan, D.~Dardari, and M.~Z. Win, ``{Position error bound for UWB
  localization in dense cluttered environments},'' \emph{IEEE transactions on
  aerospace and electronic systems}, vol.~44, no.~2, pp. 613--628, 2008.

\bibitem{shao2022reconfigurable}
X.~Shao, L.~Cheng, X.~Chen, C.~Huang, and D.~W.~K. Ng, ``Reconfigurable
  intelligent surface-aided {6G} massive access: Coupled tensor modeling and
  sparse bayesian learning,'' \emph{IEEE Transactions on Wireless
  Communications}, vol.~21, no.~12, pp. 10\,145--10\,161, 2022.

\bibitem{ETSI38_104}
\emph{Technical Specification Group Radio Access Network; Base Station (BS)
  radio transmission and reception}, European Telecommunications Standards
  Institute, 09 2021, 3GPP TS 38.104 version 15.14.0.

\bibitem{liu2017prospective}
Y.~Liu, X.~Shi, S.~He, and Z.~Shi, ``Prospective positioning architecture and
  technologies in {5G} networks,'' \emph{IEEE Network}, vol.~31, no.~6, pp.
  115--121, 2017.

\end{thebibliography}

\section*{Biographies}
\vskip -2\baselineskip plus -1fil
\begin{IEEEbiographynophoto}
{Guillermo Encinas-Lago} (M'21) received his M.Sc. degrees in Applied Physics from Universidad Autónoma de Madrid in 2013 and in Industry 4.0 from Univesidad Carlos III de Madrid in 2021, both in Spain, and a Ph.D. in Telecommunications from Université Paris-Saclay, France, in 2024, while employed at NEC Laboratories Europe in Heidelberg, Germany, under the MSCA program. He works now as a senior researcher in i2Cat, focused on reconfigurable intelligent surfaces, machine learning, localization techniques and prototyping.
\end{IEEEbiographynophoto}
\vskip -2\baselineskip plus -1fil

\begin{IEEEbiographynophoto}
{Vincenzo Sciancalepore} (M'15--SM'19) received his M.Sc. degree in Telecommunications Engineering and Telematics Engineering in 2011 and 2012, respectively, whereas in 2015, he received a double Ph.D. degree. Currently, he is a Principal Researcher at NEC Laboratories Europe, focusing his activity on reconfigurable intelligent surfaces. He is an Editor of the IEEE Transactions on Wireless Communications (since 2020) and IEEE Transactions on Communications (since 2024).
\end{IEEEbiographynophoto}
\vskip -2\baselineskip plus -1fil

\begin{IEEEbiographynophoto}{Henk Wymeersch} (S'01, M'05, SM'19, F'24) obtained the Ph.D. degree in Electrical Engineering/Applied Sciences in 2005 from Ghent University, Belgium. He is currently a Professor of Communication Systems with the Department of Electrical Engineering at Chalmers University of Technology, Sweden. Prior to joining Chalmers, he was a postdoctoral researcher from 2005 until 2009 with the Laboratory for Information and Decision Systems at the Massachusetts Institute of Technology. Prof. Wymeersch served as Associate Editor for IEEE Communication Letters (2009-2013), IEEE Transactions on Wireless Communications (since 2013), and IEEE Transactions on Communications (2016-2018) and is currently Senior Member of the IEEE Signal Processing Magazine Editorial Board.  During 2019-2021, he was an IEEE Distinguished Lecturer with the Vehicular Technology Society.  His current research interests include the convergence of communication and sensing, in a 5G and Beyond 5G context.
\end{IEEEbiographynophoto}
\vskip -2\baselineskip plus -1fil

\begin{IEEEbiographynophoto}{Marco di Renzo} (Fellow IEEE) received the Laurea (cum laude) and Ph.D. degrees in electrical engineering from the University of L’Aquila, Italy, in 2003 and 2007, respectively, and the Habilitation à Diriger des Recherches (Doctor of Science) degree from University Paris-Sud (currently Paris-Saclay University), France, in 2013. Currently, he is a CNRS Research Director (Professor) and the Head of the Intelligent Physical Communications group in the Laboratory of Signals and Systems (L2S) at Paris-Saclay University -- CNRS and CentraleSupelec, Paris, France. He is a Fellow of the IEEE, IET, EURASIP, and AAIA; an Academician of AIIA; an Ordinary Member of the European Academy of Sciences and Arts, an Ordinary Member of the Academia Europaea; an Ambassador of the European Association on Antennas and Propagation; and a Highly Cited Researcher. 

He served as the Editor-in-Chief of IEEE Communications Letters during the period 2019-2023, and he is currently serving as the Director of Journals of the IEEE Communications Society.
\end{IEEEbiographynophoto}
\vskip -2\baselineskip plus -1fil

\begin{IEEEbiographynophoto}{Xavier Costa-Perez} (M'06--SM'18) is Research Professor at ICREA, Scientific Director at the i2Cat R\&D Center and Head of 6G Networks R\&D at NEC Laboratories Europe. 
Xavier served on the Program Committee of several conferences (including IEEE ICC and INFOCOM), published at top research venues and holds several patents. 
He received his M.Sc. and Ph.D. degrees in Telecommunications from the Polytechnic University of Catalonia (UPC) in Barcelona.
\end{IEEEbiographynophoto}

\end{document}